\documentclass{article}

\usepackage{arxiv}

\usepackage[utf8]{inputenc} 
\usepackage[T1]{fontenc}    
\usepackage{hyperref}       
\usepackage{url}            
\usepackage{booktabs}       
\usepackage{amsfonts}       
\usepackage{nicefrac}       
\usepackage{microtype}      
\usepackage{lipsum}
\usepackage{graphicx}
\graphicspath{ {./images/} }
\usepackage{multirow}%
\usepackage{amsmath,amssymb,amsfonts}%
\usepackage{amsthm}%
\usepackage{mathrsfs}%
\usepackage[title]{appendix}%
\usepackage{xcolor}%
\usepackage{textcomp}%
\usepackage{manyfoot}%
\usepackage{booktabs}%
\usepackage{algorithm}%
\usepackage{algorithmicx}%
\usepackage{algpseudocode}%
\usepackage{listings}%
\usepackage{tabularx}  %
\usepackage{array}      %
\usepackage{multirow}   %
\usepackage{subfig}
\usepackage{placeins}

\newcommand{\etal}{\textit{et al.}}

\title{Dflow-SUR: Enhancing Generative Aerodynamic Inverse Design using Differentiation Throughout Flow Matching}

\author{
 Aobo Yang \\
  Department of Mechanical and Aerospace Engineering \\
  The Hong Kong University of Science and Technology \\
  Clear Water Bay, Hong Kong SAR, China \\
  \texttt{ayangae@connect.ust.hk} \\
   \And
 Zhen Wei \\
  Computer Vision Lab \\
  EPFL \\
  Rte Cantonale, Lausanne, 1015, Vaud, Switzerland \\
  \texttt{zhen.wei@epfl.ch} \\
   \And
 Rhea P. Liem \\
  Department of Aeronautics \\
  Imperial College London \\
  London, SW7 2AZ, United Kingdom \\
  \texttt{r.liem@imperial.ac.uk} \\
   \And
 Pascal Fua \\
  Computer Vision Lab \\
  EPFL \\
  Rte Cantonale, Lausanne, 1015, Vaud, Switzerland \\
  \texttt{pascal.fua@epfl.ch} \\
}

\begin{document}
\maketitle
\begin{abstract}
Generative inverse design requires the consideration of physical constraints in exploring new designs to make generation reliable and accurate. We observe that state-of-the-art energy-based approaches exhibit an \textit{asynchronous phenomenon} in which optimization of the physical loss is throttled by flow matching inference. To address this issue, we introduce \textit{Dflow-SUR}, a differentiation strategy that decouples physical loss optimization from flow matching inference. \textit{Dflow-SUR} lowers the physical loss by four orders of magnitude compared with the strongest energy-based baseline while trimming wall-clock time by 74\% on airfoil case and boosts the mean lift-to-drag ratio by 11.8\% over traditional Latin-hypercube sampling on wing case. In addition to accuracy and speed, \textit{Dflow-SUR} delivers three practical benefits: (i) superior guidance controllability, (ii) reduced surrogate uncertainty, and (iii) robustness to hyper-parameter tuning. Collectively, these results underscore \textit{Dflow-SUR}’s promise as a scalable, high-fidelity framework for generative aerodynamic design.
\end{abstract}


\section{Introduction}
\label{sect:introduction}

Generative aerodynamic inverse design is a data-driven approach that leverages generative models to propose high-performance designs. It has emerged as an alternative to traditional discriminative design, which  performs an optimization to find a single optimal solution~\cite{li2021machine, Lyu.AESCTE.2024} and relies on a surrogate model for performance estimation. This requires a high-quality initial guess, such as conceptual design, which can be difficult to obtain. In contrast, generative design operates without such dependency and enables a broader exploration of the design space. It does so by learning the implicit distribution of valid aerodynamic shapes from existing data, enabling probabilistic sampling of feasible configurations. Some examples in deep generative models include variational autoencoders (VAEs)~\cite{KingmaWelling2014}, generative adversarial networks (GANs)~\cite{GoodfellowEtAl2014}, and flow-based generative models, which rely on \textit{normalizing flows} to define expressive probability distributions from which data are generated~\cite{Papamakarios2021.FlowModels}. With the advancement of flow models for the latter category, such as diffusion models~\cite{ho2020denoising, song2021score} and flow matching~\cite{lipman2022flow}, the denoising paradigm has enabled more controllable and high-fidelity generative processes. Furthermore, physics-based guidance can be provided during shape generation. This steers the generative process towards generating geometrically valid and high-performance aerodynamic shapes. These may feature non-intuitive innovations beyond traditional parameter limits while still meeting performance goals.



\begin{figure}[htbp]
    \centering
    \includegraphics[width=1.0\linewidth]{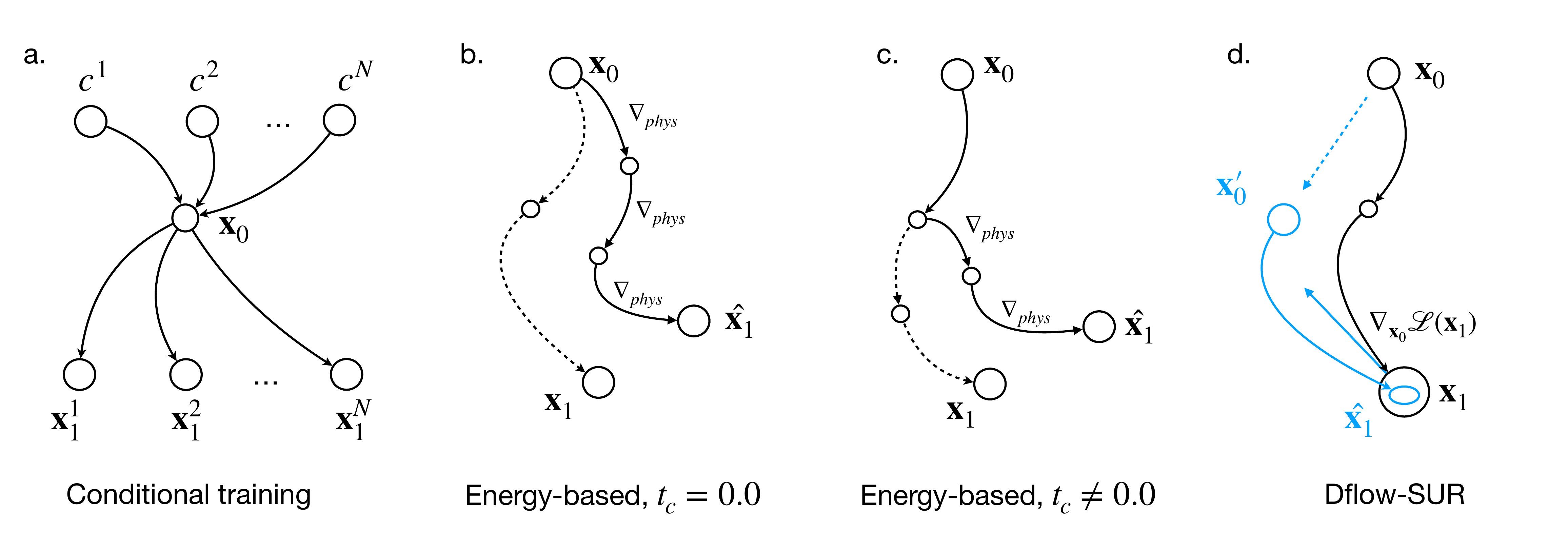}
    \caption{Four physical guidance generation strategies: a. conditional guidance during flow matching training; b and c represent energy-based approach with different $t_c$ setting; d is the newly developed \textit{Dflow-SUR}. Strategies shown in b, c, and d involve physical guidance during flow matching inference.}
    \label{fig:fmStrategy}
\end{figure}

Physics can be incorporated into generative models in two ways: via conditional training or through inference-time guidance, depending on when the physics guidance is introduced. Figure~\ref{fig:fmStrategy} shows four representative physical guidance strategies of generative design.  
Figure~\ref{fig:fmStrategy}a depicts the conditional training strategy, where the physical loss $\mathcal{L}_{\mathrm{phys}}$ is combined with the flow matching loss during model training. This strategy has been adopted in previous studies~\cite{lu2023contrastive,yang2024data}. The remaining strategies apply physical guidance during inference: the energy-based approach is shown in Figures~\ref{fig:fmStrategy}b and \ref{fig:fmStrategy}c (to be discussed in Section~\ref{subsubsect:energy}) and the proposed \textit{Dflow-SUR} approach is illustrated in Figure~\ref{fig:fmStrategy}d (to be detailed in Section~\ref{subsubsect:diff}).


In conditional training, the physical loss $\mathcal{L}_{\mathrm{phys}}$ is typically combined with the flow-matching loss into a single composite objective, serving as a conditional signal alongside the design $\mathbf{x}$ under which the neural network jointly learns the velocity field parameterization. In the context of aerodynamic inverse design, conditional training has been applied to solve multipoint~\cite{Lin2025MultiPoint} and multifidelity problems~\cite{yang2024data}. Lin~\etal~\cite{Lin2025MultiPoint} implemented a classifier-free conditioning by randomly dropping and concatenating performance targets as inputs during diffusion training so the model learns to generate airfoil shapes both with and without explicit condition guidance, eliminating the need for separate classifier networks. Yang~\etal~\cite{yang2024data} trained a conditional diffusion model by optimizing a score network on noisy shapes given performance targets and a value function network via contrastive learning of predicted target values to guide sampling toward the desired aerodynamic performance. However, this approach is fundamentally constrained by its training paradigm. First, it lacks flexibility, as any modification to design objectives or constraints necessitates retraining the model. Second, accurately modeling both the design space and the underlying physics jointly demands substantially more data, increasing the burden on data collection and model complexity. Third, the conditioning mechanism is limited to low-dimensional settings, where only a few scalar values can be used as auxiliary inputs to the network.


Currently, the prevailing consensus is to introduce physical guidance during the inference phase of flow model. In this framework, a pre-trained, unconditional flow model handles solely geometry sampling, whereas the physical constraint is incorporated at certain inference time. The physics loss is typically cast as an energy model~\cite{lecun2006tutorial}, giving rise to the energy-based approach (Figures~\ref{fig:fmStrategy}b and~\ref{fig:fmStrategy}c). In this approach, physical losses are formulated as an energy equation and its gradients are used to guide the flow model during inference. In the engineering design domain, energy-based generative models constitute a new trend, gaining attention for their ability to incorporate physical constraints into the generation process. Wu~\etal~\cite{wu2024compositional} developed the Compositional Inverse Design with Diffusion Models (CinDM) method, which reframes inverse design as energy minimization via diffusion, compositing multiple diffusion-based energy functions over overlapping subsets of variables. The composite strategy ensures that designs remain in-distribution locally while generalizing to multi-body designs. In the context of topology optimization, TopoDiff~\cite{maze2023diffusion} injects physical information during intermediate inference stages, leveraging low-uncertainty data for physical generation to help control sample quality. To further reduce inference costs, the diffusion optimization models (DOM) developed by Giannone~\etal~\cite{giannone2023aligning} adopt a trajectory alignment regularizer, which tightly couples diffusion sampling with physics-based optimization steps and kernel relaxations. 

\begin{figure}
    \centering
    \includegraphics[width=0.8\linewidth]{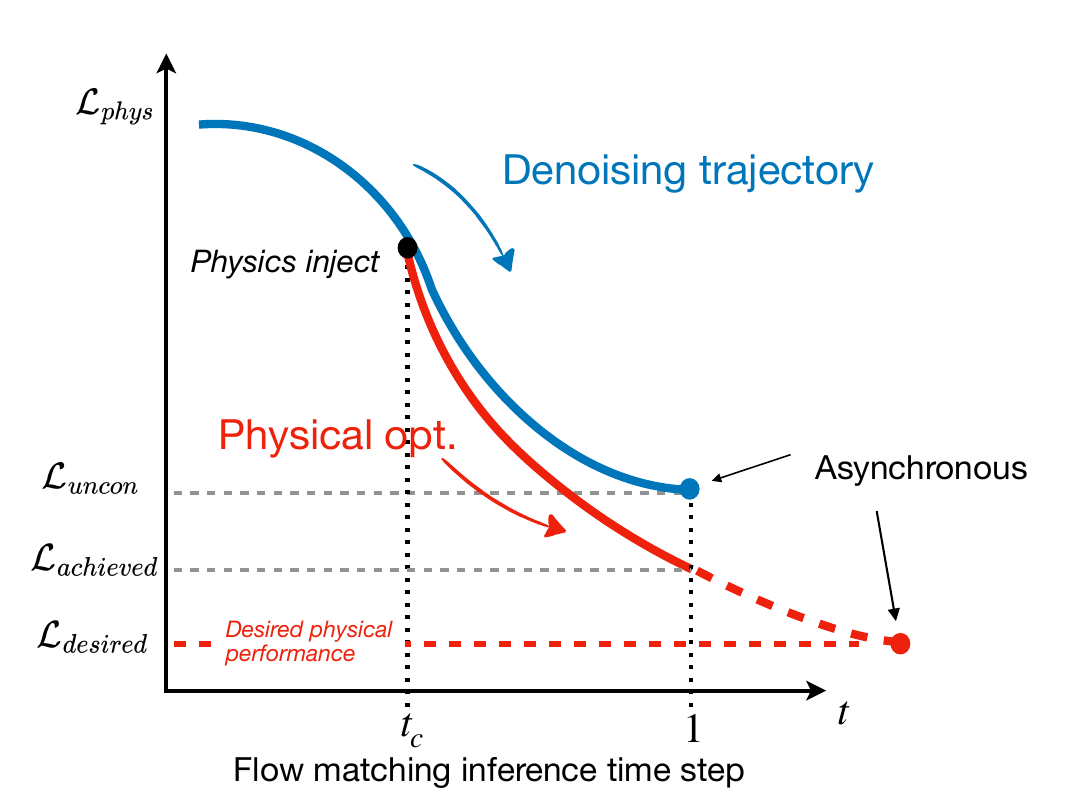}
    \caption{Asynchronous dynamics between flow-matching denoising and physical loss optimization. $\mathcal{L}_{\text{uncon}}$ denotes the final physical loss from unconditional generation, $\mathcal{L}_{\text{achieved}}$ represents the final physical loss under energy-based guidance, and $\mathcal{L}_{\text{desired}}$ indicates the target physical constraint imposed on the generative model.}
    \label{fig:asynchronous}
\end{figure}

Despite the demonstrated effectiveness mentioned above, the energy-based approach is constrained by two key issues. The core reason is that surrogate models are typically trained on simulation data from clean, idealized geometries, and the limited training data often result in low robustness when faced with more complex or even noisy, perturbed designs. Under such conditions, two main problems emerge: high predictive uncertainty from the surrogate model and strong sensitivity to hyperparameters, such as the injection time $t_c$ and the energy coefficient $\lambda$. All these issues result in a phenomenon we newly find, termed \textit{asynchronous phenomenon}, leading to failed generation. As illustrated in Figure~\ref{fig:asynchronous}, the finite inference budget from energy-based approach inherently limits physical-loss optimization. An overly short inference time schedule may lead to insufficient physical loss optimization, while an excessively long schedule increases the computational cost of inference. We refer to the inability to synchronize physical-loss optimization with the generative inference process---i.e., when both processes cannot finish at the same time---as an \textit{asynchronous phenomenon}. As a result, one must manually calibrate multiple hyperparameters (such as energy coefficient $\lambda$, inference time steps $T$ and injection time $t_c$, etc.) to ensure that the generated samples follow a reasonable distribution and achieve the desired reduction in physical loss within the allotted inference budget.

To address this issue, we build upon recent advances in differentiable flow matching \textit{D-Flow}~\cite{ben2024dflow} to propose \textit{Dflow-SUR}. It is a decoupled framework that separates inference from physical guidance by updating the generative trajectory using gradients evaluated only on the final generated sample. 
The key contribution of this paper is the demonstration that coupling \textit{D-Flow} with physical surrogate modeling enhances controllability and mitigates uncertainty in surrogate predictions, thereby establishing a significantly more effective generative inverse design framework than previous approaches. 

More specifically, we exhibit
%
%
through extensive experiments in both conditional training and energy‐based inference, four main strengths of \textit{Dflow-SUR}:
\begin{itemize}
    \item \textbf{Superior guidance controllability.} By decoupling the inference dynamics from the physical‐loss gradient, \textit{Dflow-SUR} completely avoids gradient conflicts and thereby improves overall model accuracy.
    \item \textbf{Low surrogate uncertainty evaluation.} In contrast to other methods whose gradients suffer from large surrogate‐model uncertainty at early denoising steps, \textit{Dflow-SUR} confines generation to the data manifold throughout, effectively controlling uncertainty.
    \item \textbf{Hyperparameter robustness.} Whilst existing approaches demand extensive manual tuning of guidance strength to achieve good results, \textit{Dflow-SUR} delivers high-quality designs without any manual adjustment of guidance hyperparameters.
    \item \textbf{Fast and accurate generation.} The decoupled design of \textit{Dflow-SUR} significantly reduces the number of inference time steps while independently optimizing the physical loss, leading to substantial improvements in both computational efficiency and physical accuracy.
\end{itemize}

This paper is organized as follows. In Section~\ref{sect:relatedWork}, we review the related works. Section~\ref{sect:methodology} then presents the methodology used in this study, followed by case studies using 2D airfoil and 3D wing presented in Section~\ref{sect:exp}. Finally,  we conclude the key findings of this work in Section~\ref{sect:conculsion}.

\section{Related works}
\label{sect:relatedWork}
The realization of \textit{Dflow-SUR} requires three key components: design representation, physical guidance with a surrogate model, and generative inverse design. We review the related works around these three aspects, which are presented below.

\subsection{Design representation}
Geometry parameterization, as a design representation approach, maps complex geometry into a low-dimensional latent space, enabling efficient learning and optimization while preserving essential features. Control point approaches such as non-uniform rational B-spline (NURBS)~\cite{dassault_nurbs,zhao2024automated} and free-form deformation (FFD)~\cite{sederberg1986ffd,hajdik2023pygeo} have been shown to be suitable and efficient for design optimization. However, during Design of Experiments (DoE) sampling, control point samples often lie outside the reasonable design space, which disrupts geometric continuity. This shortcoming in turn complicates learning-based optimization by obscuring underlying patterns or constraints. Previous studies have employed data‐subspace techniques—such as singular value decomposition (SVD)~\cite{Masters2017}, proper orthogonal decomposition (POD)~\cite{WU2019632pod}, and class-shape transformation (CST)~\cite{Kulfan2008}—to render design representations more controllable. For three-dimensional geometries, global‐feature parameterization methods—such as compact modal parameterization~\cite{li2021adjoint}—have been adopted to capture geometric modes. More advanced approaches such as latent representation method~\cite{wei2023latent,wei2023automatic} and neural-network-based method~\cite{wei2024deepgeo,wei2025automated} have subsequently been introduced to bolster representation robustness and afford greater flexibility by expanding the design’s degrees of freedom. In this study, we employ the CST for airfoil parameterization and the compact modal parameterization scheme for three-dimensional wing geometries, 
which have been shown effective in previous aerodynamic shape optimization studies~\cite{Lyu.AESCTE.2024, Yang.JoA.2025}. Furthermore, the effectiveness of their derivative computation plugins has also been demonstrated~\cite{li2021adjoint,hajdik2023pygeo}.

\subsection{Surrogate-assisted design}
A key enabler in physics-informed generative design is a surrogate model that serves as a fast approximation model to emulate the behavior of expensive physics evaluations~\cite{Liem.AST.2015, li2021machine, Paiva2010}. The low-cost gradient information provided by the surrogate model is valuable for rapid conceptual design phases where computational efficiency is critical. The surrogate model construction can be based on Gaussian processes~\cite{Liem.AST.2015, Bouhlel2019102662,hwang201874}, multi-layer perceptron~\cite{li2021data}, geodesic convolutional neural networks~\cite{baque2018geodesic}, and transfer learning~\cite{yang2025rapid} combined with different formations of neural networks, to name a few. By leveraging deep neural networks to enhance the surrogate model, we can push its representation of the physical state toward a higher fidelity level and greater dimensionality on solving complex aerodynamic scenarios, such as wing shape shock mitigation~\cite{li2021data}, buffet-onset constraint modeling~\cite{Li2022buffet}, multipoint performance optimization~\cite{Lin2025MultiPoint}, and transonic drag reduction \cite{zhang2021multi}. When integrated within a generative modeling, surrogate models can be used during the generation process to impose physics constraints or evaluate candidates on-the-fly without resorting to full simulations. This integration has enabled the exploration of new airfoil designs~\cite{li2019fast,chen2020airfoil} and topology structures~\cite{giannone2023aligning}. In this context, surrogate models offer the advantage of enabling rapid approximation of design variables within the existing data space through high-dimensional interpolation. However, a critical limitation arises in their handling of out-of-distribution (OOD) data, where uncertainty propagates significantly~\cite{owen2017compare}.

\subsection{Generative aerodynamic design}
The main purpose of using generative models is to explore new design spaces. Previous approaches such as variational autoencoders (VAEs)~\cite{KingmaWelling2014} and generative adversarial networks (GANs)~\cite{GoodfellowEtAl2014} have been used to do shape parameterization (e.g., B\'ezier GAN~\cite{chen2020airfoil}, PaDGAN~\cite{chen2021padgan}), geometric filtering~\cite{li2020efficient}, and accommodate constrained design candidates~\cite{achour2020development,wang2022inverse,lei2021deep}. In recent years, diffusion models~\cite{ho2020denoising, song2021score} have emerged as powerful deep generative models with proven performance in computer vision, natural language processing, and modeling of different types of data. This method
employs a progressive denoising probabilistic framework that ensures stable training and exact likelihood evaluation, thereby mitigating mode collapse and producing samples with higher fidelity than GANs and VAEs~\cite{ho2020denoising,song2021score}. With \textit{DiffAirfoil}, Wei~\etal~\cite{wei2024diffairfoil} demonstrated diffusion model’s superiority over GANs under data-scarce conditions. In the context of aerodynamic design, our previous study demonstrated the effectiveness of conditional diffusion model as a geometry sampling approach to generate high-fidelity aerodynamic performance wings~\cite{yang2024data}. Several works have also applied conditional diffusion model on multipoint~\cite{Lin2025MultiPoint} and multi-body~\cite{wu2024compositional} settings in similar design problems. Another generative modeling method is flow matching~\cite{ben2024dflow}, which directly learns a continuous velocity field to map base distribution samples to the target distribution. This method is adopted in this work and will be further discussed in Section~\ref{subsect:flowBackground}.

\section{Methodology}
\label{sect:methodology}

In this section, we first introduce the flow matching background in Section~\ref{subsect:flowBackground}. We then introduce the physical guidance strategy in the flow model in Section~\ref{subsect:physics}, which leads to the discussion on the proposed \textit{Dflow-SUR} strategy.

\subsection{Flow matching}
\label{subsect:flowBackground}
Flow matching~\cite{lipman2022flow} is an approach for training continuous normalizing flows (CNFs) based on regressing vector fields of fixed conditional probability paths. A CNF is designed to learn complex data distributions by parameterizing a time-dependent velocity field $u_t^{\theta}:\mathbb{R}^d \times [0,1]\rightarrow \mathbb{R}^d$ between Gaussian distributions $\mathcal{N}(0, I)$ and a target distribution $X$ to generate the final state $\mathbf{x}_1$ by solving the ODE, 
\begin{equation}
    \frac{d \mathbf{x}_t}{d t}
    = u_t^{\theta}\bigl(\mathbf{x}_t, t\bigr), 
    \quad \mathbf{x}_0 \sim \mathcal{N}(0, I), 
    \quad \mathbf{x}_1 = \Phi_{[0,1]}\bigl(\mathbf{x}_0\bigr),
\end{equation}
where $u_t^{\theta}\bigl(\mathbf{x}_t, t\bigr)$ denotes the learned velocity field, $\mathbf{x}_0$ is the initial noise, and $\theta$ represents the neural network parameters that define it. The typical unconditional flow matching training process is summarized in Algorithm~\ref{alg:flowMatching}, where $\eta$ is the learning rate for gradient descent on $\theta$.

\begin{algorithm}
    \caption{Flow Matching Training}
    \label{alg:flow_matching}
    \begin{algorithmic}
        \Require Data distribution $X$, noise distribution $\mathcal{N}(0,I)$
        \State Initialize vector field parameters $\theta$ of $u_t^\theta(x)$
        \For{each training iteration}
          \State Sample $\mathbf{x}_0\sim \mathcal{N}(0,I)$, $\mathbf{x}_1 \in X$, and $t \sim U(0,1)$
          \State Compute interpolated point
            \[
              \mathbf{x}_t = (1-t)\mathbf{x}_0 + t\mathbf{x}_1
            \]
          \State Compute target velocity
            \[
              \dot{\mathbf{x}}_t = \frac{\mathbf{x}_1-\mathbf{x}_0}{1-t}
            \]
          \State Compute regression loss
            \[
              \mathcal{L}(\theta)=\bigl\|u_t^\theta(\mathbf{x}_t)-\dot{\mathbf{x}}_t\bigr\|^2
            \]
          \State Update parameters:
            \[
              \theta \;\gets\;\theta - \eta\,\nabla_\theta \mathcal{L}(\theta)
            \]
        \EndFor
    \end{algorithmic}
    \label{alg:flowMatching}
\end{algorithm}

\subsection{Physics injection strategy}
\label{subsect:physics}
When injecting physical loss $\mathcal{L}_{\mathrm{phys}}$ optimization into a pre-trained flow matching velocity field inference process, there are two iteration parameters that serve distinct levers: flow matching inference-time discretization parameter $T$ and the physical loss optimization iteration parameter $K$. Their roles are described briefly below:
\begin{itemize}
    \item Inference-time discretization $T$ determines the numerical fidelity of the flow map from initial noise $\mathbf{x}_0$ to final state $\mathbf{x}_1$. During the inference phase, the continuous time interval $[0, 1]$ is partitioned into $T$ segments, with a single step size defined as $\Delta t = \frac{1}{T}$. 
    \item The number of physical loss optimization iterations $K$ denotes the required iterations for enabling $\mathbf{x}_1$ to approximate the desired physical performance,  where $\Delta k$ represents one optimization iteration step.
\end{itemize}
Both parameters can be independently adjusted; however, their step sizes are also regulated by the physical loss introduction mechanism and optimization strategy. 

\subsubsection{Energy-based strategy}
\label{subsubsect:energy}
Physical-based conditional generative approaches typically adopt the energy-based method (EBM)~\cite{lecun2006tutorial,teh2003energy}, which is a probabilistic framework that defines a distribution through an energy function. In this strategy, lower energy states correspond to more probable configurations,
\begin{equation}
p(\textbf{x} \mid \textbf{c}) \propto p(\textbf{x}) \cdot \mathrm{e}^{-\lambda \cdot \mathcal{E}(\textbf{x})},
\label{eqn:energyEqn}
\end{equation}
where physical constraints on design $\textbf{x}$ are formed in the energy equation $\mathcal{E}(\textbf{x})$, controlled by the energy coefficient $\lambda$. Originating from thermal theory, $\lambda$ regulates the trade-off between data exploitation and exploration in the generative model. Specifically, smaller $\lambda$ values prioritize leveraging the current data sample distribution $p(\textbf{x})$, whereas larger $\lambda$ values emphasize exploration guided by the energy function $\mathcal{E}(\textbf{x})$.

Given a pre-trained flow matching velocity field $u_t^{\theta}\bigl(\mathbf{x}_t, t\bigr)$ parameterized by $\theta$, the physics-guided generation process can thus be expressed as 
\begin{equation}
    \frac{d \mathbf{x}_t}{d t} = u_t^{\theta}\bigl(\mathbf{x}_t, t\bigr) - \lambda \nabla_{\mathbf{x}_t} \mathcal{E}\left(\mathbf{x}_t\right).
\label{eqn:energy}
\end{equation} 
Based on the distinct stages at which the energy function $\mathcal{E}\left(\mathbf{x}_t\right)$ is introduced into the flow matching time interval $t \in [0,1]$, we categorize the strategies into two energy-based cases, as illustrated in Figures~\ref{fig:fmStrategy}b and \ref{fig:fmStrategy}c. Specifically, if the cutoff time of $\mathcal{E}\left(\mathbf{x}_t\right)$ is defined as $t_c$, the two strategies correspond to the scenarios where $t_c = 0$ (Figure~\ref{fig:fmStrategy}b) and $t_c \neq 0$ (Figure~\ref{fig:fmStrategy}c), respectively. When $t_c = 0$, the strategies integrate physical guidance from the initial denoising step ($t=0$), ensuring continuous enforcement of physical constraints throughout the entire generation trajectory. The $\mathbf{x}$ generation trajectory is updated with Equation~\ref{eqn:energy}.

However, due to the high noise level in the initial generation phase, the optimization process is significantly influenced by the gradient uncertainty of the surrogate model over noisy samples, which may degrade generation performance. To mitigate this, the inter-drifting strategy is proposed~\cite{maze2023diffusion,giannone2023aligning}, in that the generative model first operates in an unconditional manner until time $t=t_c$ to ensure partial denoising of the design, after which physical guidance is incrementally injected to refine the trajectory. The $\mathbf{x}$ generation trajectory is updated with
\begin{equation}
\frac{d\mathbf{x}_t}{dt}
=
\begin{cases}
u_t^{\theta}\bigl(\mathbf{x}_t, t\bigr), 
& t < t_c, \\[6pt]
u_t^{\theta}\bigl(\mathbf{x}_t, t\bigr)
- \lambda\,\nabla_{\mathbf{x}_t}\mathcal{E}(\mathbf{x}_t), 
& t \ge t_c.
\end{cases}
\end{equation}
During this procedure, the number of iterations $K$ available for minimizing the physical loss becomes intrinsically tied to the inference schedule. Once physics‐based guidance is activated at time $t_c$, the remaining inference horizon $t - t_c$ directly limits the optimization depth, such that
\begin{equation}
    K = \frac{t - t_c}{\Delta t}, \quad \Delta k = \Delta t.
\end{equation}
This mechanism determines that the physical loss optimization inherently depends on the fidelity of the inference-time discretizations. In other words, the later the physics term is introduced, the fewer gradient-descent steps can be performed on the physical loss within the fixed inference budget.

The intermediate-injection strategy ($t_c \neq 0$) reduces the uncertainty of surrogate model estimations to the generated samples and ensures the gradient quality. However, the finite inference budget inherently limits physical-loss optimization. An inference schedule that is too brief precludes effective loss minimization, whereas an excessively extended schedule incurs unnecessary computational cost. As introduced in Section~\ref{sect:introduction}, the \textit{asynchronous phenomenon} refers to the inability to synchronize physical-loss optimization with the generative inference process when both processes cannot finish at the same time. This discrepancy arises because the energy-based optimization (via $\mathcal{E}(\mathbf{x})$) and the flow-matching denoising (via $p(\mathbf{x})$) operate with distinct objectives and temporal dependencies. Detailed procedure of the energy-based approach can be found in Algorithm~\ref{alg:energyAlg} presented in Appendix~\ref{sect:App_energyBasedAlg}.

In summary, asynchronous behavior in the energy‐based framework arises because physical‐loss optimization (guided by $\nabla\mathcal{E}(\mathbf{x})$) and flow‐matching inference (driven by $u_{t}(\mathbf{x})$) pursue different objectives on different time scales. To synchronize them, one must manually tune two hyperparameters: the energy coefficient $\lambda$, which controls the strength of the physical constraint, and the total number of inference steps $T$, which determines the duration of flow matching. Careful adjustment of $\lambda$ and $T$ is essential to balance physical plausibility against generative fidelity.

\subsubsection{Dflow-SUR}
\label{subsubsect:diff}
In this section, we present the proposed \textit{Dflow-SUR} method. Using this approach, we evaluate $\mathcal{L}(\mathbf{x}_1)$ and back-propagate the loss gradient through the flow ODE to update the initial noise $\mathbf{x}_0$. \textit{D-Flow}~\cite{ben2024dflow}, which is used as the base method to develop \textit{Dflow-SUR}, is a strategy that enables a fully differentiable inference process by tracing the influence of the evaluation on final state $\mathbf{x}_1$ throughout to initial noise $\mathbf{x}_0$. In particular, we recast controllable generation as a source‐point optimization that benefits both from high‐quality gradients---since $\mathbf{x}_1$ has already been denoised by the model---and from an implicit data‐manifold projection. In this case, the ODE’s Jacobian filters out off‐manifold components and thus injects the model’s learned prior into each update, jointly ensuring fidelity to the target objective and consistency with the design space of interest.
 
Here, we combine the differential throughout flow matching strategy with a surrogate model, denoted as \textit{Dflow-SUR}. Given a trained flow matching vector field $u^{\theta}(\mathbf{x})$, \textit{Dflow-SUR} enables a fully differentiable inference process by tracing the influence of the initial noise $\mathbf{x}_0$ through to the final state $\mathbf{x}_1$, which sets it apart from conventional sampling-based methods. This allows us to directly optimize the initial noise to steer the generated design toward desired outcomes. Let \(\mathrm{SUR}_{\phi}(\mathbf{x})\) denote the surrogate physics evaluator, parameterized by weights \(\phi\). The loss function can then be expressed as
\begin{equation}
    \mathcal{L}(\mathbf{x})=\|\mathrm{SUR}_{\phi}(\mathbf{x})-\mathbf{y}\|^2,
\end{equation}
where $\mathbf{y}$ is the desired physical performance. We examine the dynamics of the final output $\mathbf{x}_1$ under optimization. Specifically, we consider updating the initial noise $\mathbf{x}_0$ via a gradient descent step:
\begin{equation}
    \mathbf{x}_0^\tau = \mathbf{x}_0 - \tau \nabla_{\mathbf{x}_0} \mathcal{L}(\mathbf{x}_1),
\end{equation}
where the gradient \(\nabla_{\mathbf{x}_0} \mathcal{L}(\mathbf{x}_1)\) is computed through the chain rule,
\begin{equation}
    \nabla_{\mathbf{x}_0} \mathcal{L}(\mathbf{x}_1) = D_{\mathbf{x}_0} \mathbf{x}_1^\top \nabla_{\mathbf{x}_1} \mathcal{L}(\mathbf{x}_1).
\end{equation}
This formulation enables end-to-end differentiation from the design objective back to the initial noise input. This process is illustrated in Figure~\ref{fig:fmStrategy}d and the corresponding pseudocode is shown in Algorithm~\ref{alg:dflowSUR}. As outlined in Algorithm~\ref{alg:dflowSUR}, differentiating through the flow matching begins by sampling an initial noise $\mathbf{x}_0$. The gradient $\nabla_{\mathbf{x}_0} \mathcal{L}(\mathbf{x}_1)$ is then computed to iteratively optimize $\mathbf{x}_0$, steering the generated sample $\mathbf{x}_1$ toward the desired constraints. Since the generation path depends on the initial $\mathbf{x}_0$, achieving diversity in outputs requires multiple optimization runs with distinct noise initializations.

\begin{algorithm}[H]
\caption{\textit{Dflow-SUR}}
\begin{algorithmic}[1]
\Require A learned vector field $u_{t}^{\theta}(\mathbf{x})$, surrogate model $\mathrm{SUR}_{\phi}(\mathbf{x})$, desired physical performance $\mathbf{y}$
\State Initialize guess $\mathbf{x}_0$
\For{$k = 1$ to $K-1$}
    \State Generate flow matching by solving an ODE
    \[
        \mathbf{x}^{k}(1) \leftarrow \text{solve}(\mathbf{x}_0^{k}, u_t)
    \]
    \State Compute gradient by chain rule
    \[
    \nabla_{\mathbf{x}_0} \mathcal{L}(\mathbf{x}_1) = D_{\mathbf{x}_0} \mathbf{x}_1^\top \nabla_{\mathbf{x}_1} \mathcal{L}(\mathbf{x}_1)
    \]
    \State Optimize initial noise
    \[
    \mathbf{x}_0^{k+1} \leftarrow \mathbf{x}_0^{k} - \tau \nabla_{\mathbf{x}_0} \mathcal{L}(\mathbf{x}^{k}_1)
    \]
\EndFor
\end{algorithmic}
\label{alg:dflowSUR}
\end{algorithm}

\section{Numerical experiments}
\label{sect:exp}
In this section, we implement the strategies described in Section~\ref{sect:methodology} on two case studies, namely the 2D airfoil inverse design (Section~\ref{sect:AirfoilCase}) and 3D wing inverse design (Section~\ref{sect:wingCase}).

\subsection{2D airfoil inverse design case}
\label{sect:AirfoilCase}
In this particular case, the generative inverse design process is required to propose airfoil candidates that satisfy the lift coefficient $C_{L} = 0.7$ when the Mach number $M=0.2$, angle of attack $\alpha=2^\circ$, and Reynolds number $Re = 1\times10^{6}$. We use \texttt{NeuralFoil}\footnote{\texttt{NeuralFoil} repository: \url{https://github.com/peterdsharpe/NeuralFoil} (last accessed on 22 July 2025)} as surrogate model for rapid airfoil aerodynamic analysis. \texttt{NeuralFoil} is implemented as a hybrid of analytical models and neural networks trained on tens of millions of \texttt{XFoil}\footnote{XFoil webpage: \url{https://web.mit.edu/drela/Public/web/xfoil/} (last accessed on 22 July 2025)} runs. \texttt{NeuralFoil} supports auto-differentiation and has been proven effective on various gradient-based design optimization cases~\cite{aerosandbox_phd_thesis}. The CST method~\cite{Kulfan2008} is used for the airfoil shape parameterization. The airfoil geometry is controlled by a total of 16 parameters, with eight shape coefficients assigned to the upper surface and the other eight to the lower surface. 

\subsubsection{Guidance controllability}
\label{subsect:physicalloss}
Here, we investigate the guidance controllability; we first investigate the $\mathcal{L}_{\mathrm{phys}}$ curve and precision achieved by the four strategies shown in Figure~\ref{fig:fmStrategy} (Section~\ref{subsect:physics}). All quantitative results of model performance are provided in Table~\ref{tab:modelPerformanceResults} in Appendix~\ref{subsect:modelPerformance}. The discussion below is focused on the results obtained using the energy-based approach.

\begin{figure}[htbp]
    \centering
    \includegraphics[width=0.8\linewidth]{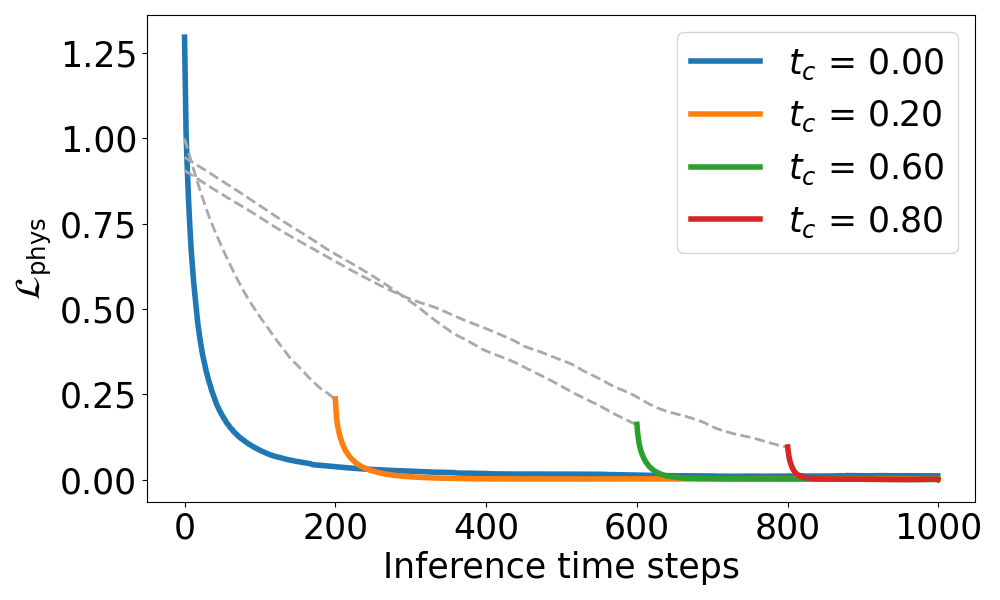}
    \caption{The mean loss curves fo energy-based approaches, when $t_c = 0.0, 0.2, 0.6, 0.8$, $T=1000$ (gray dashed lines represent \textit{pseudo‐loss} curves).}
    \label{fig:mean_loss_curve_1k}
\end{figure}

For the energy-based approach, we generate $200$ samples with $\lambda = 10$ under different physics injection cutoff times ($t_c = 0.0, 0.2, 0.6, 0.8$) and total inference time steps ($T = 200, 1000, 2000$). Figure~\ref{fig:mean_loss_curve_1k} shows the loss decay curve when $T=1000$. 
We compute the surrogate model’s \textit{pseudo‐loss} ($\mathcal{L}_{\mathrm{phys}}$ during unconditional generation) for different $t_c$ values, which are shown as gray dashed lines. These lines illustrate the $\mathcal{L}_{\mathrm{phys}}$ decay behavior under the energy-based approach's inference-physics coupling optimization process. Unconditional generation over certain inference time steps provides a good initialization by achieving a lower initial $\mathcal{L}_{\mathrm{phys}}$.

\begin{figure}[htbp]
    \centering
    \subfloat[Physical loss ($\mathcal{L}_{\mathrm{phys}}$ of final generated airfoils)\label{subfig:phys_loss}]{%
        \includegraphics[width=0.8\textwidth]{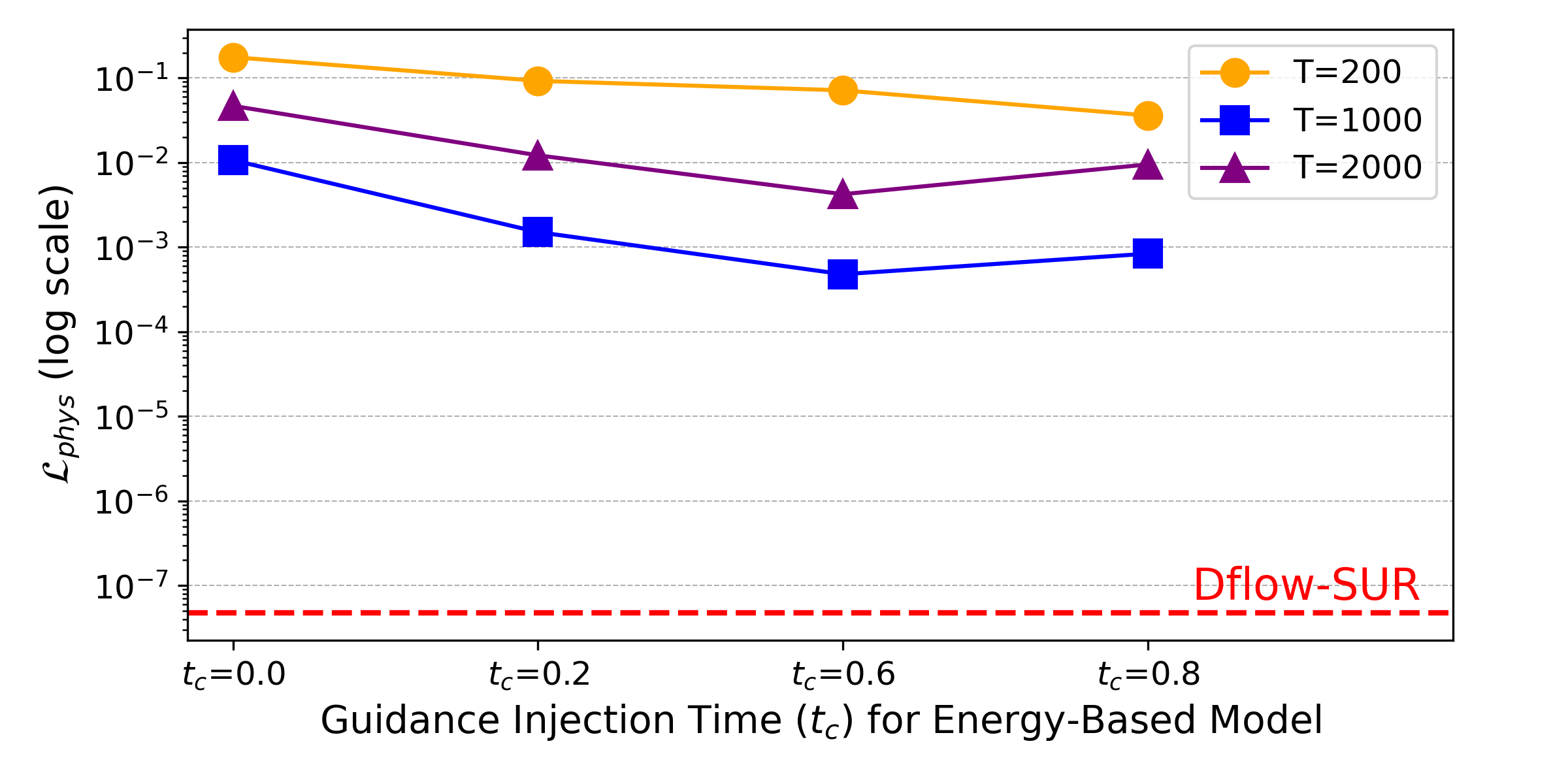}%
    }\\[1em]
    \subfloat[Accuracy of lift coefficient ($C_{L}$) ($C_{L}$ of generated airfoils, with targeted $C_{L}=0.7$)\label{subfig:cl_acc}]{%
        \includegraphics[width=0.8\textwidth]{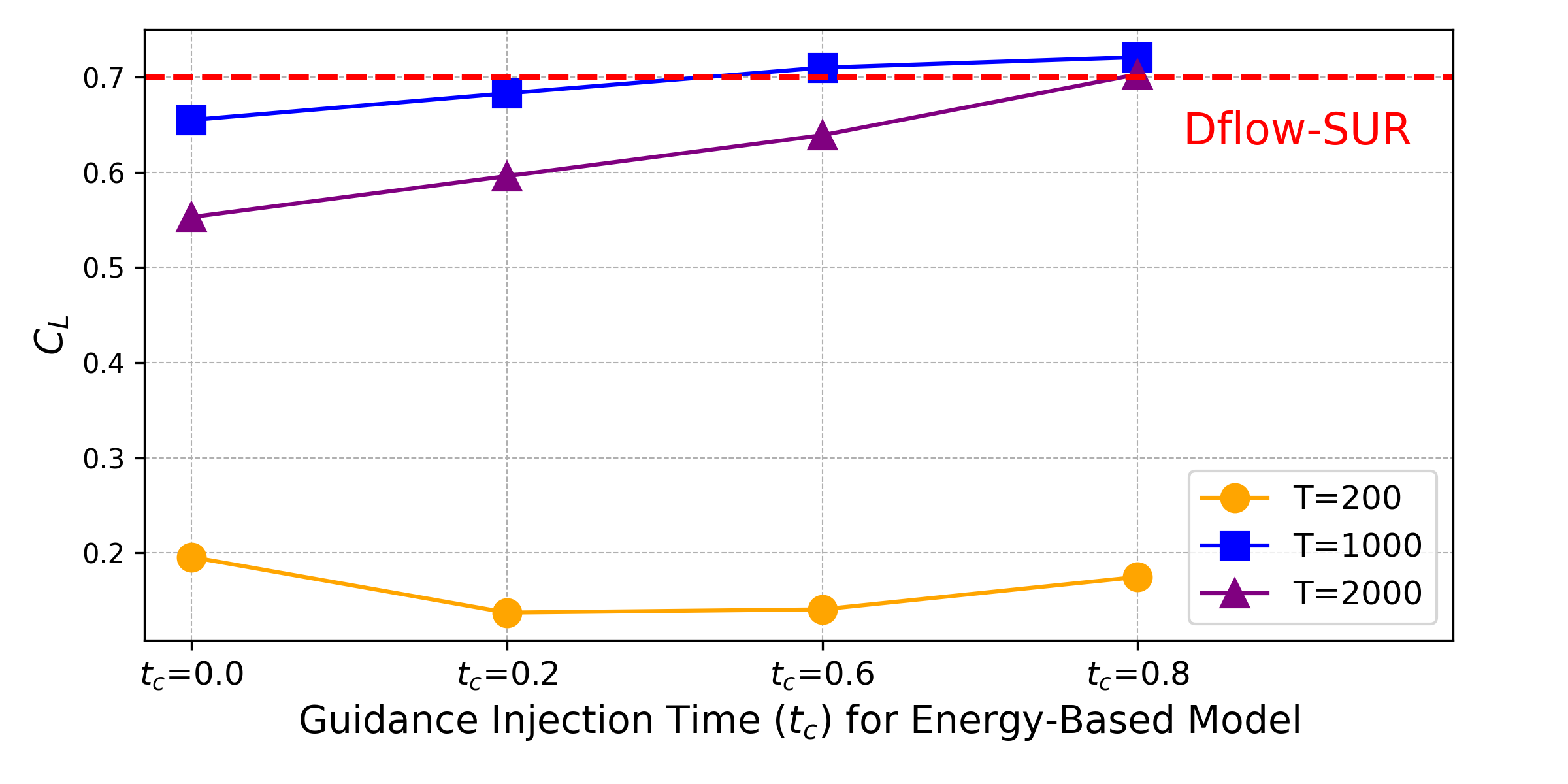}%
    }\\[1em]
    \subfloat[Inference time\label{subfig:inf_time}]{%
        \includegraphics[width=0.8\textwidth]{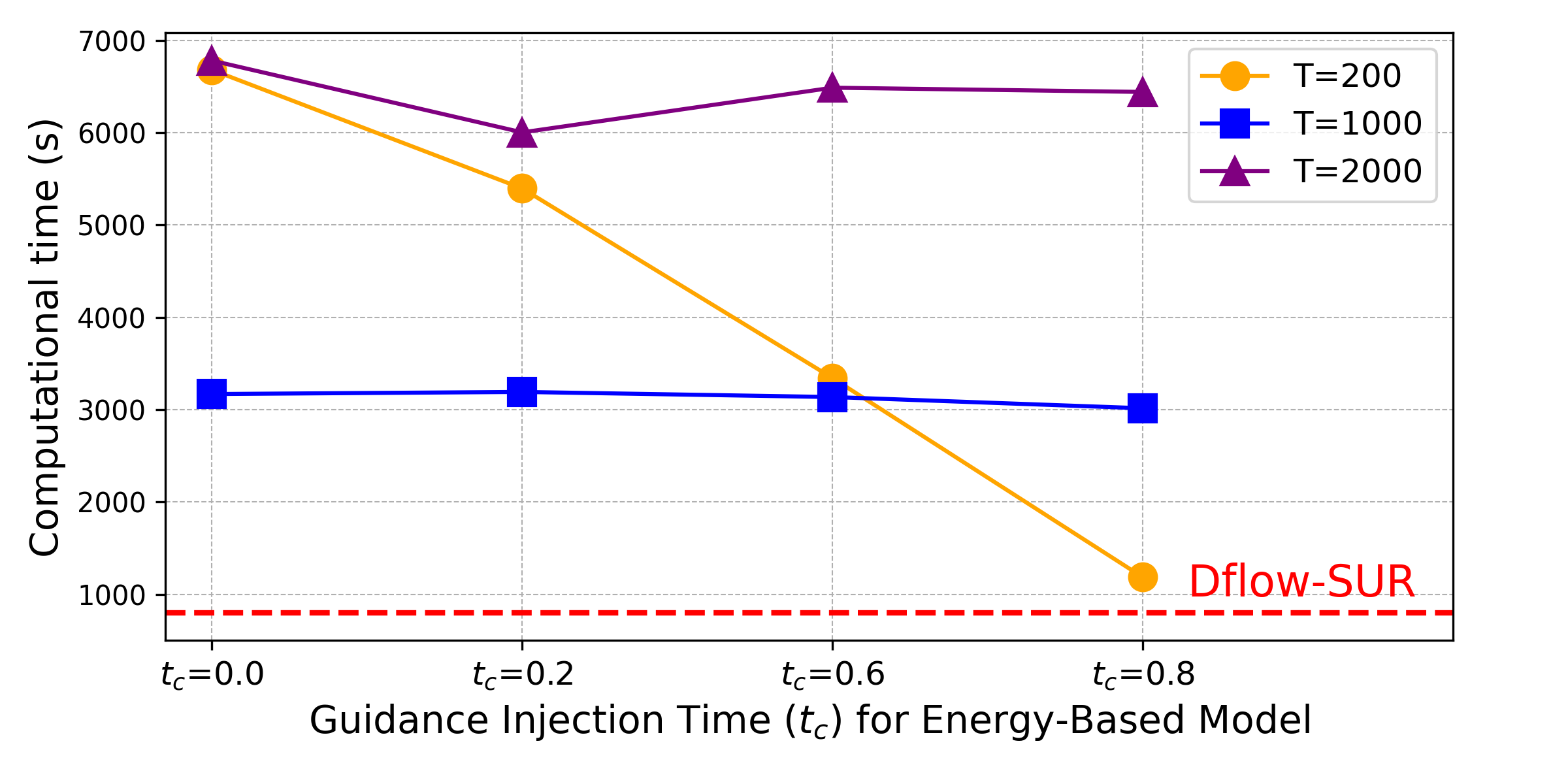}%
    }
    \caption{Performance comparison between the energy-based approach and \textit{Dflow‐SUR} in three metrics.}
    \label{fig:modelPerformanceCompare}
\end{figure}

Next, we apply \textit{Dflow‐SUR} to 200 samples. Specifically, we draw 200 random initial guesses, execute the \textit{Dflow‐SUR} procedure on each sample, and record their trajectories. We then compare the performance of both the energy‐based approach and \textit{Dflow‐SUR} in terms of $\mathcal{L}_{\mathrm{phys}}$, $C_{L}$ accuracy, and inference time, which are shown in Figure~\ref{fig:modelPerformanceCompare}. Overall, compared to the energy-based method, \textit{Dflow-SUR} meets the aerodynamic constraint of $C_L = 0.7$ while driving the $\mathcal{L}_{\mathrm{phys}}$ down from $10^{-3}$ to $10^{-8}$ and achieving the shortest inference time. 

This improvement arises chiefly from \textit{Dflow-SUR}’s decoupling of the inference process from the $\mathcal{L}_{\mathrm{phys}}$ optimization, which enhances gradient‐utilization efficiency compared to the tightly coupled mechanism of energy-based methods. Decoupling is effective because the inference and physics-loss objectives exhibit \textit{gradient collision} when optimized jointly. To illustrate this phenomenon in detail, we perform the following experiments.




Referring to the work by Wang~\etal~\cite{wang2025gradient}, we denote the gradients of the velocity field $g^v_k$ and of the physical‐loss term $g^p_k$ at each inference time step $t_k$, with respect to the generated sample $\mathbf{x}_k$, by
\begin{equation}
    g^v_k \;=\;\nabla_{\mathbf{x}}\,f_\theta(\mathbf{x}_k,t_k)
    \quad\text{and}\quad
    g^p_k \;=\;\nabla_{\mathbf{x}}\,\mathcal{L}_{\mathrm{ ys}}(\mathbf{x}_k). 
\end{equation}
We then define their alignment score as
\begin{equation}
    \mathrm{Align}\bigl(g^v_k,\,g^p_k\bigr)
    \;=\;
    \frac{\bigl\langle g^v_k,\,g^p_k\bigr\rangle}
         {\bigl\lVert g^v_k\bigr\rVert\,\bigl\lVert g^p_k\bigr\rVert}\,.
\end{equation}
In particular, the alignment score takes the value \(+1\) when the two gradients are perfectly co‐directional, \(-1\) when they are exactly opposite, and vanishes (or is near zero) when they are orthogonal or cancel each other out.

\begin{figure}[htbp]
    \centering
    \includegraphics[width=0.8\linewidth]{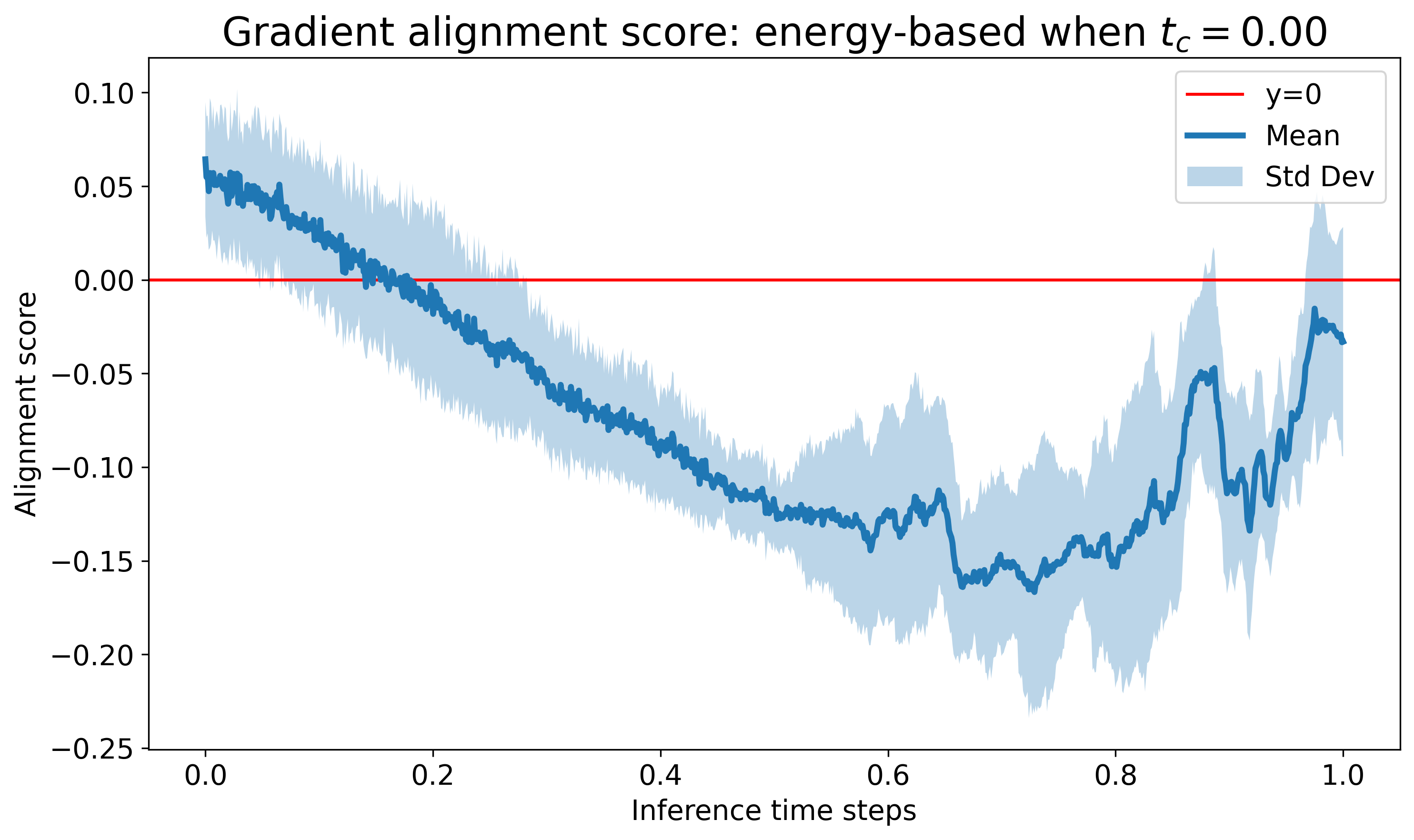}
    \caption{The gradient alignment score of energy-based approach when $t_c = 0.0$, $T=1000$.}
    \label{fig:alignmentScore}
\end{figure}

We plot the alignment score of the energy-based approach with $t_c = 0.0$ and $T=1000$ in Figure~\ref{fig:alignmentScore}. It can be clearly observed that the alignment score remains negative for most of the time, with slight small positive values at the initial stage of inference. This phenomenon indicates that the directions of $g^v$ and $g^p$ are predominantly opposed during optimization. We term this behavior \textit{gradient collision}. The persistence of \textit{gradient collision} suggests that flow matching inference and physical loss minimization do not align in their optimization directions for the majority of iterations. Notably, the energy-based coupling between inference and physical loss exacerbates this conflict. In contrast, \textit{Dflow-SUR} decouples the inference process from physical loss optimization, allowing them to operate independently. This decoupling mechanism is a key contributor to \textit{Dflow-SUR}’s superior accuracy compared to coupled energy-based methods.

In summary, by decoupling inference from physical‐loss optimization, \textit{Dflow‐SUR} achieves gradient‐collision‐free guidance and enhanced guidance controllability, leading to several orders-of-magnitude reduction in physical loss.

\subsubsection{Surrogate model uncertainty quantification}
\label{subsect:uncertainty}
In this section, we discuss and quantify the uncertainty of the surrogate model in estimating generated samples along the trajectory. Following a previous study by Gal and Ghahramani~\cite{gal2016dropout}, we employ Monte Carlo dropout of neurons of the surrogate model deep net with $1\%$ neuron deactivation rate. For each generated sample, we perform $20$ independent forward passes to collect $20$ predictions. We take the standard deviation of $20$ surrogate model estimations per sample for uncertainty quantification (UQ) purposes. Thus, we define UQ for each generated sample $\mathbf{x}^k$ as 
\begin{equation}
\mathrm{UQ}(\mathbf{x}^k)=\sigma(\mathbf{x}^k)=\sqrt{\frac{1}{N-1} \sum_{i=1}^N\left(f^{(i)}(\mathbf{x}^k)-\mu(\mathbf{x}^k)\right)^2},
\end{equation}
where $N=20$ is the number of stochastic forward passes, $f^{(i)}(\mathbf{x})$ is the prediction of pass $i$, $\mu(\mathbf{x})=\frac{1}{T} \sum_{i=1}^T f^{(i)}(\mathbf{x})$ is the prediction pass mean.

\begin{figure}[htbp]
    \centering
    \includegraphics[width=1.0\linewidth]{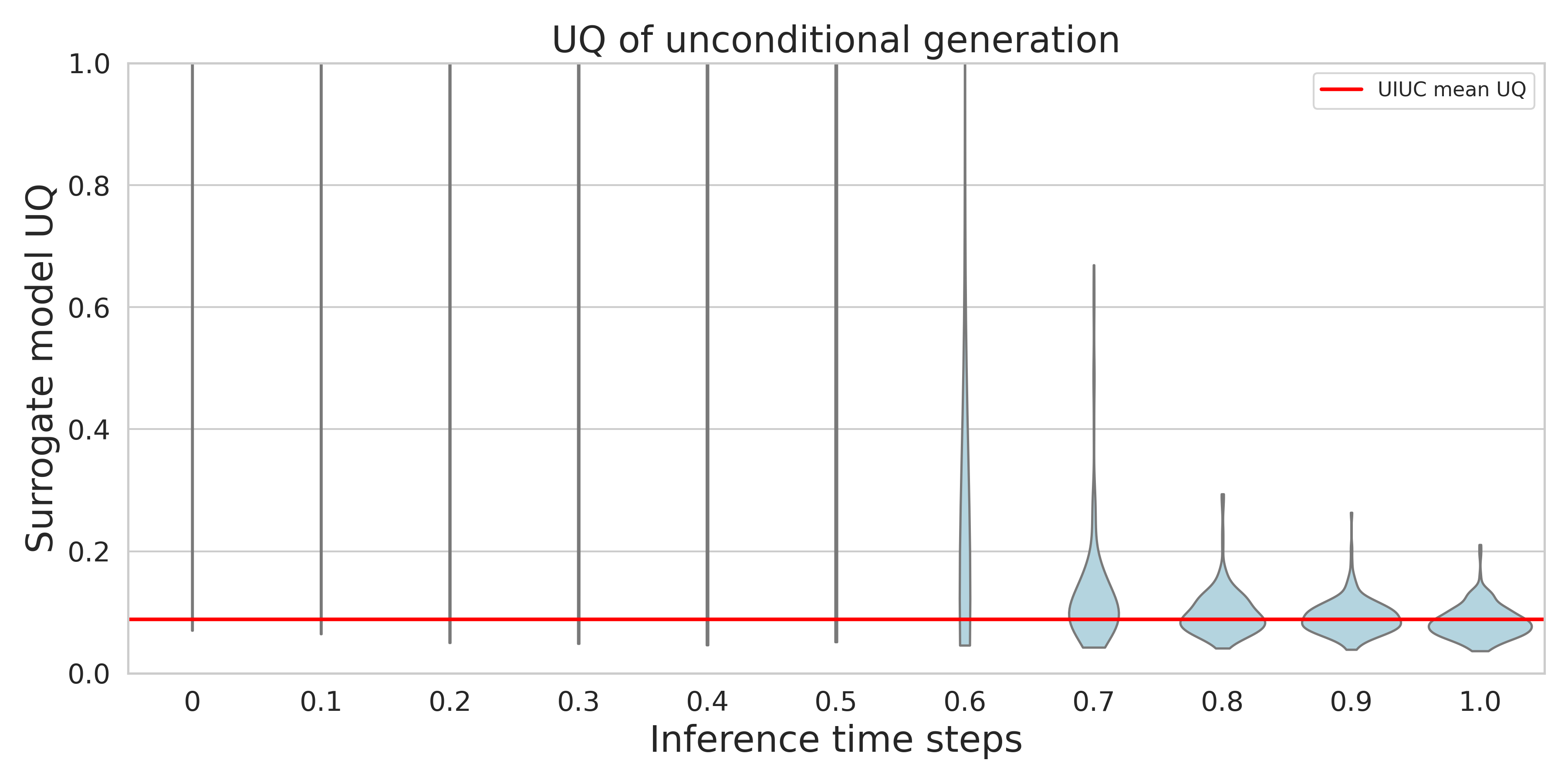}
    \caption{The normalized violin plot illustrates the uncertainty (UQ) of unconditionally generated samples along the denoising trajectory, with the red line indicating the mean UQ of UIUC airfoils.}
    \label{fig:uqUnconditional}
\end{figure}

To assess the overall uncertainty profile of the generated designs, we plot a violin diagram of UQ values for the full batch of 200 samples. Figure~\ref{fig:uqUnconditional} shows the statistical visualization of the distribution of surrogate model UQ for generated samples along an unconditional generation trajectory (i.e., at different time steps). The red line represents the mean UQ of the surrogate model predictions for the UIUC training dataset\footnote{UIUC Airfoil Coordinates Database: \url{https://m-selig.ae.illinois.edu/ads.html} (last accessed on 23 July 2025). The database is established by the Applied Aerodynamics Group at the University of Illinois Urbana-Champaign.}. Observations reveal that during most of the early stages of the generated trajectory, the generated samples notably deviate from the mean line, indicating substantial uncertainty. Consequently, the gradients of the $\mathcal{L}_{\mathrm{phys}}$ connected to the surrogate model become inaccurate in this phase. Combined with other observations  illustrating four physics injection strategies for energy-based methods (shown in Figure~\ref{fig:uqEnergy} in Appendix~\ref{sect:AppendixUncertainty}), it remains challenging to identify an optimal physics injection time $t_c$ that balances model uncertainty while ensuring thorough optimization of the $\mathcal{L}_{\mathrm{phys}}$.

\begin{figure}[htbp]
    \centering
    \includegraphics[width=1.0\linewidth]{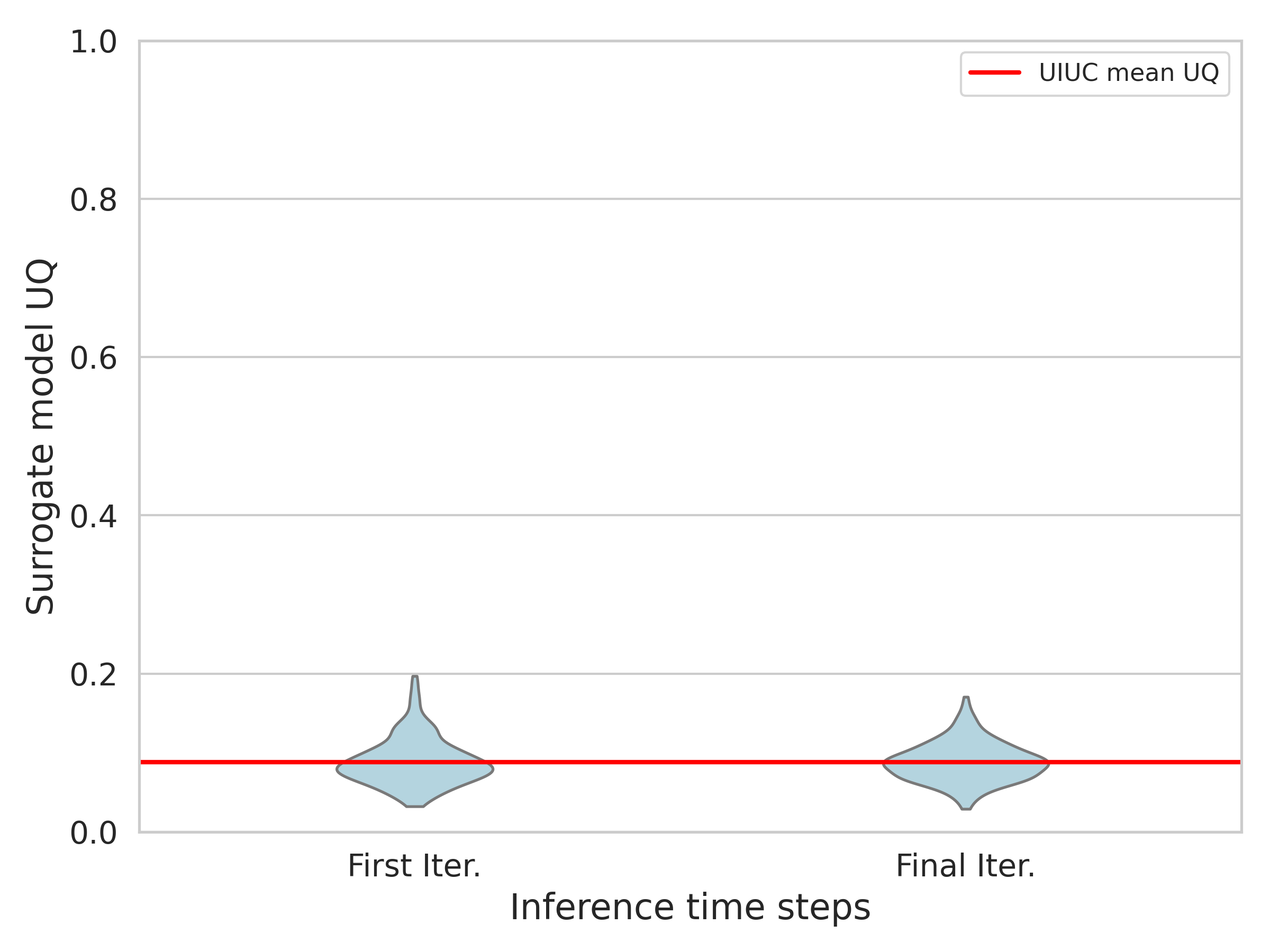}
    \caption{The UQ of \textit{Dflow-SUR} generates samples along the time scheduling trajectory (the red line represents the mean of UIUC airfoil UQs).}
    \label{fig:uqDflowSUR}
\end{figure}

Next, we show the UQ of the surrogate model for the first and final iterations generated by \textit{Dflow-SUR} in Figure~\ref{fig:uqDflowSUR}. The sample distributions of these two datasets are concentrated around the mean value (red line), thereby ensuring the stability of the surrogate model's gradients. This is formally guaranteed by Theorem 4.2 of D-Flow \cite{ben2024dflow}, which ensures that the method inherently generates samples confined to the data manifold. Under the Affine Gaussian Probability Path (AGPP) (i.e., Equation 12 in Ref.~\cite{lipman2022flow}) assumption, they show that the Jacobian of the mapping from the noise input $\mathbf{x}_0$ to the output $\mathbf{x}_1$ is a time-ordered exponential of local covariance matrices,
\begin{equation}
D_{\mathbf{x}_0} \mathbf{x}_1=\sigma_1 \mathcal{T} \exp \left[\int_0^1 \gamma_t \operatorname{Var}_{1 \mid t}(\mathbf{x}(t)) d t\right],
\label{eqn:manifold}
\end{equation}
where $\mathcal{T} \exp [\cdot]$ stands for a time-ordered exponential and $\gamma_t$ is defined as $\gamma_t=\frac{1}{2} \frac{d}{d t} \frac{\alpha_t^2}{\sigma_t^2}$, in which $\alpha_t$ is the mean-scaling coefficient that defines AGPP and is used to interpolate between noise and data. Equation~\ref{eqn:manifold} indicates that each infinitesimal update projects the gradient onto the principal directions of data variance (i.e., the data manifold). Discretizing this ODE with \(N\) uniform Euler steps of size \(h = 1/N\) yields the Jacobian of \(\mathbf{x}_1\) with respect to \(\mathbf{x}_0\):
\begin{equation}
D_{\mathbf{x}_0}\,\mathbf{x}_1
= \prod_{m=0}^{N-1}
  \Bigl(
    (1 + h\,a_{m h})\,I
    \;+\;
    h\,\gamma_{m h}\,\operatorname{Var}_{1\mid m h}\bigl(x_{m h}\bigr)
  \Bigr).
\end{equation}
Consequently, the optimization trajectory remains confined to the data distribution by iteratively applying the above covariance‐based projections.

From the above explanation, we can surmise that the energy‐based approach generates inference trajectories with large uncertainty, which undermines the surrogate model’s ability to provide reliable gradient guidance for physical‐loss optimization. In contrast, \textit{Dflow‐SUR} constrains generation along the data manifold, keeping uncertainty close to the model's mean value and avoiding manifold drifting.

\subsubsection{Sensitivity of guidance strength}
\label{subsect:diversity}
The energy-based approach requires quite a substantial manual hyperparameter tuning, which makes the process very sensitive. Table~\ref{tab:energy_diff_lambda} summarizes the experiments of energy-based approach generation with different energy coefficient $\lambda$ settings.  
\begin{table}[htbp]
    \centering
    \caption{$\mathcal{L}_{\mathrm{phys}}$ and $C_L$ for energy-based approach (when $t_c = 0.6$, T=1000) with different $\lambda$ settings.}
    \label{tab:energy_diff_lambda}
    \begin{tabular}{c c c}
        \toprule
        Energy Coefficient $\lambda$ & $\mathcal{L}_{\mathrm{phys}}$ & $C_L$ \\
        \midrule
        $1$      & $2.032 \times 10^{-2}$  & 0.565  \\
        $10$     &  $4.91 \times 10^{-4}$ & 0.706  \\
        $100$    &  $1.765 \times 10^{-4}$& 0.711\\
        $1000$   & $1.02 \times 10^{-5}$ & 0.703 \\
        $10000$  & Fail & Fail\\
        \bottomrule
    \end{tabular}
\end{table}
As stated in Section~\ref{subsubsect:energy}, $\lambda$ controls the energy-based approach's data exploitation and exploration. Setting this parameter either too low or too high can severely degrade the generative performance and may even cause the model to fail.

\begin{figure}[htbp] 
    \centering
    \mbox{
        \subfloat[Energy-based: $\lambda = 0.1$, $t_c=0.0$. \label{subfig:airfoils_1}]{\includegraphics[width=0.4\linewidth]{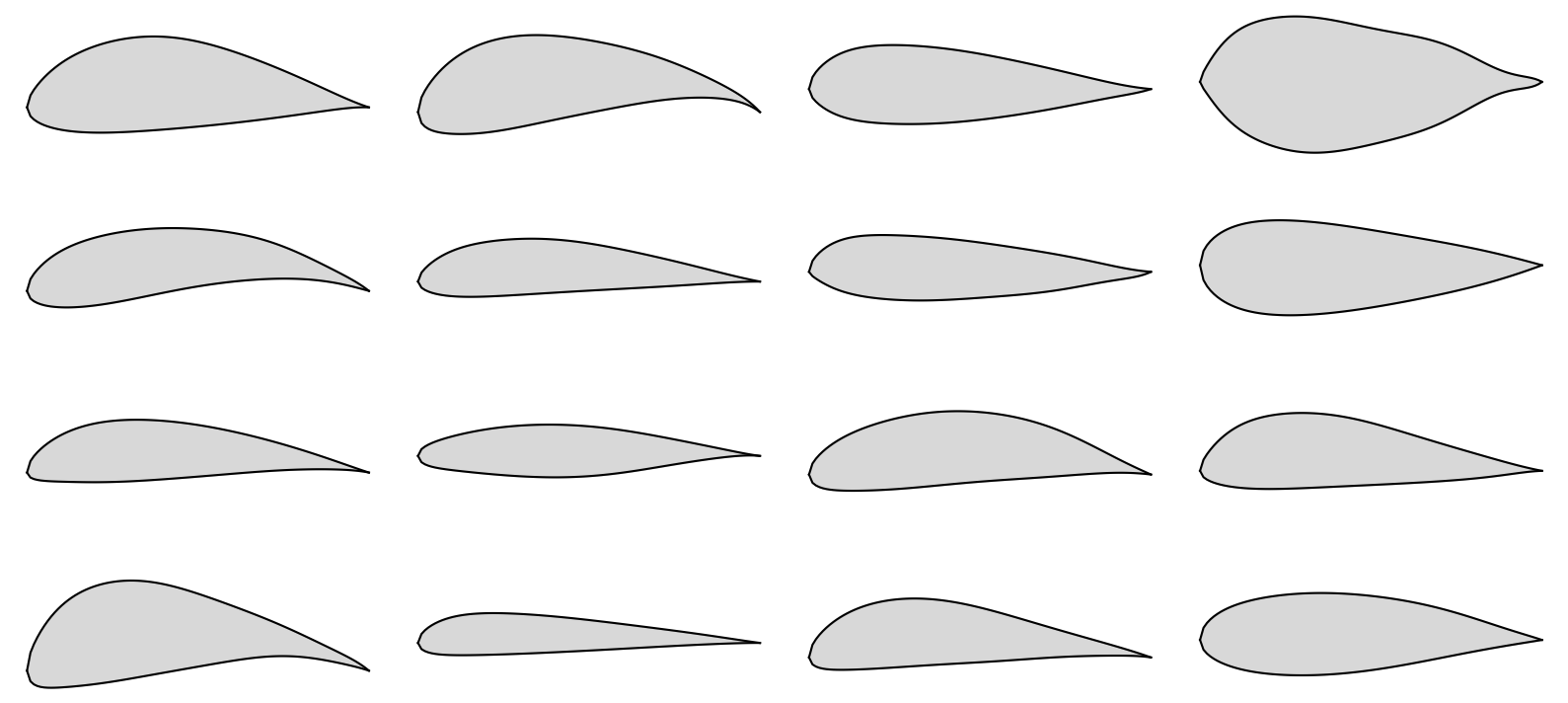}}
        \subfloat[Energy-based: $\lambda = 0.1$, $t_c=0.6$. \label{subfig:airfoils_2}]{\includegraphics[width=0.4\linewidth]{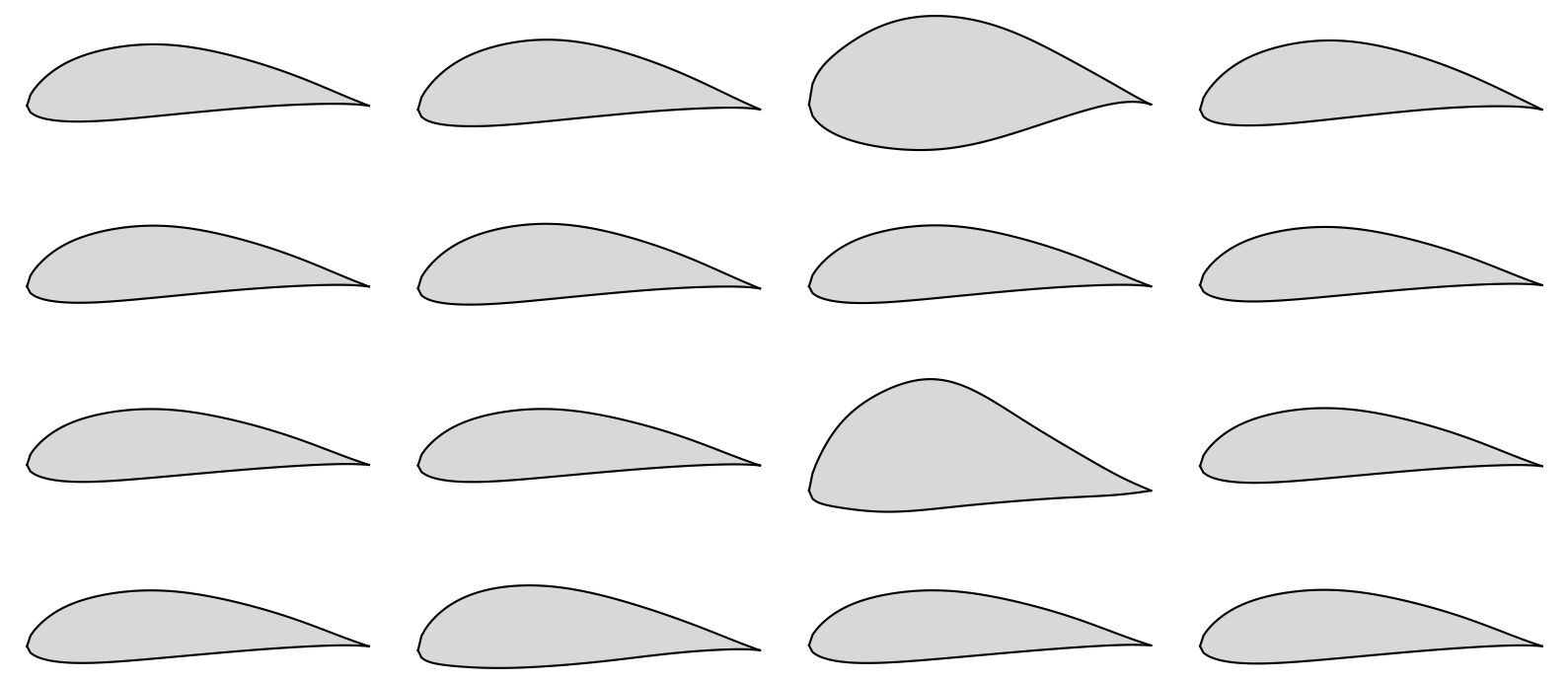}}
    }\\

    \mbox{
        \subfloat[Energy-based: $\lambda = 10$, $t_c=0.0$. \label{subfig:airfoils_2}]{\includegraphics[width=0.4\linewidth]{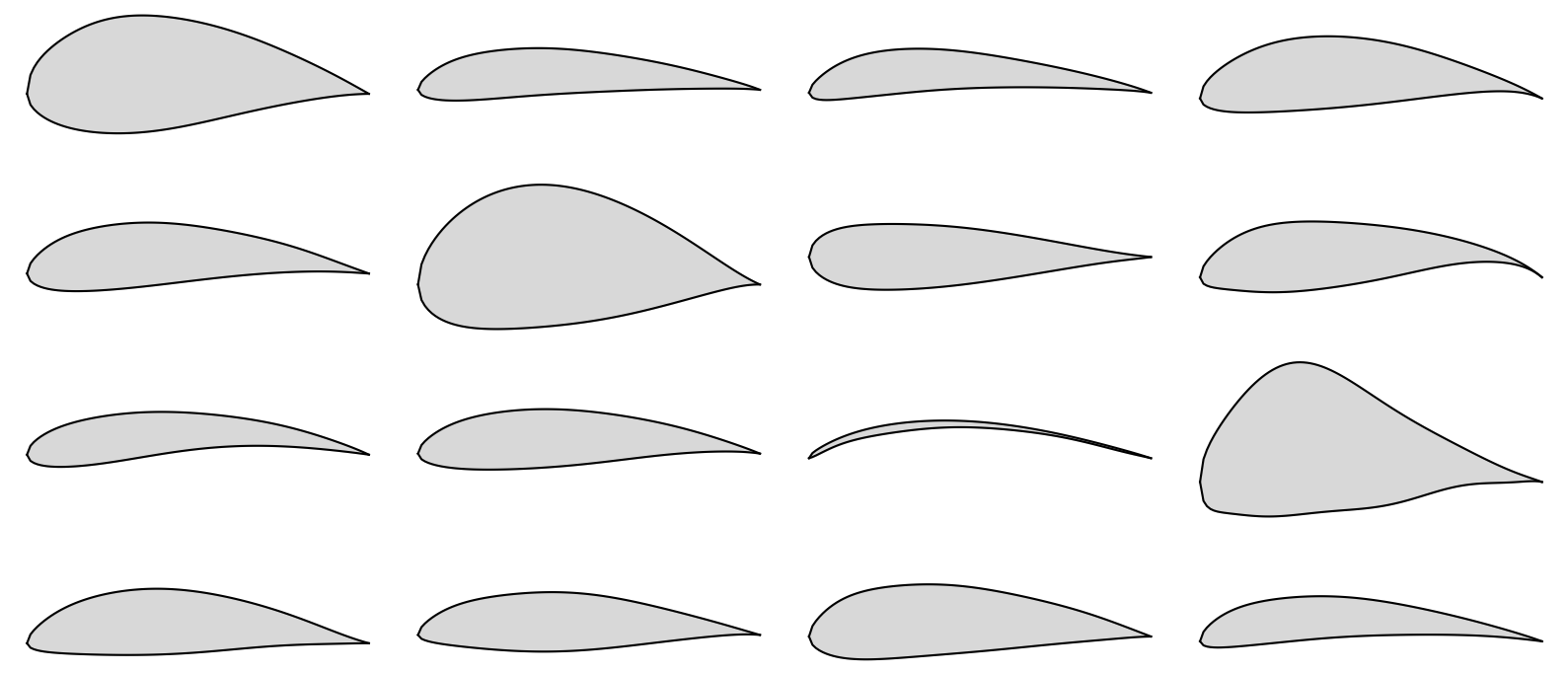}}
        \subfloat[Energy-based: $\lambda = 10$, $t_c=0.6$. \label{subfig:airfoils_4}]{\includegraphics[width=0.4\linewidth]{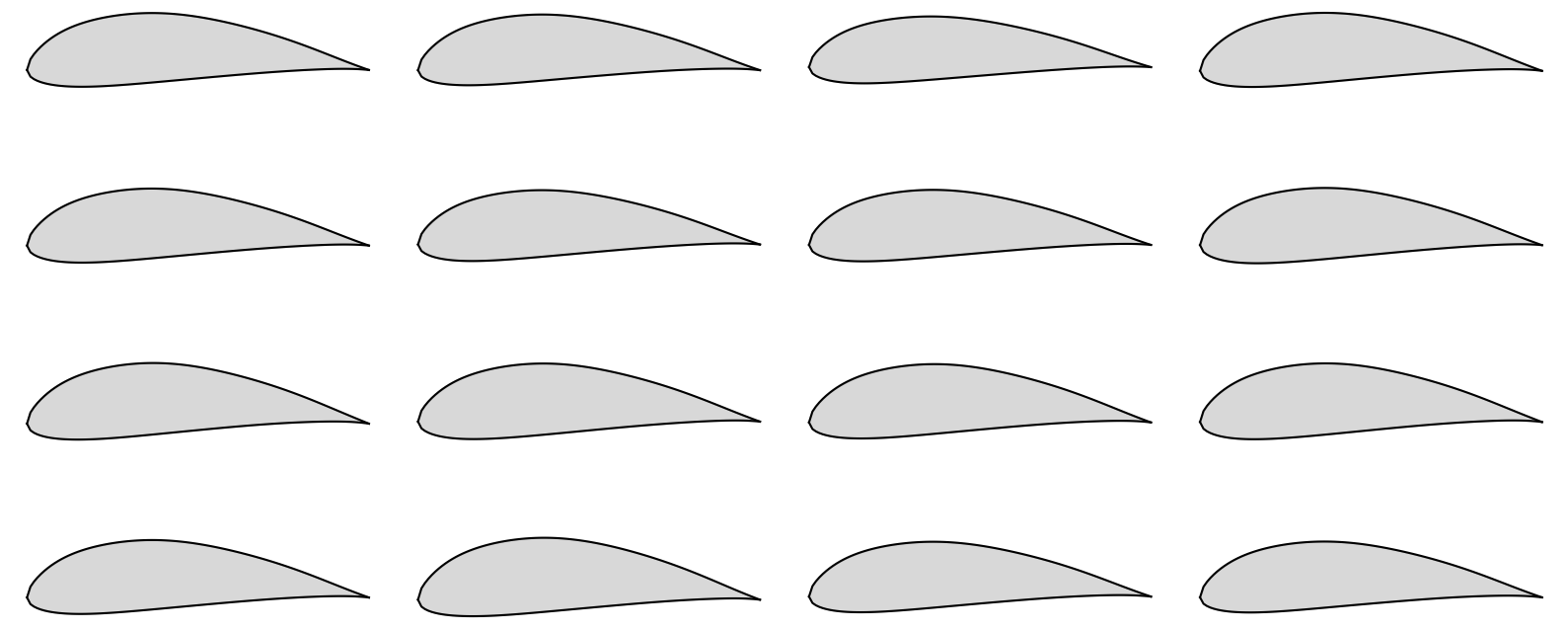}}
    }\\

    \mbox{

        \subfloat[Energy-based: $\lambda = 100$, $t_c=0.0$. \label{subfig:airfoils_3}]{\includegraphics[width=0.4\linewidth]{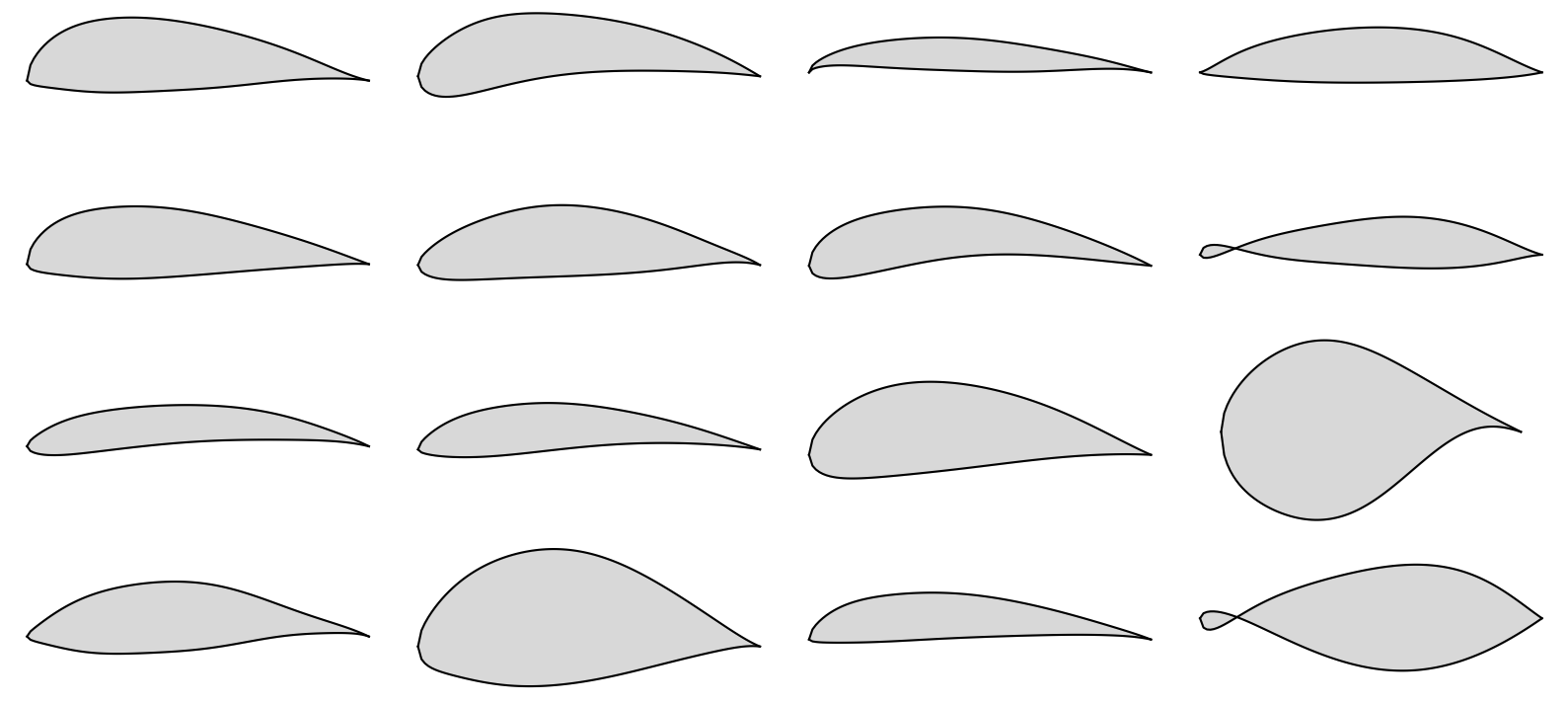}}
        \subfloat[\textit{Dflow-SUR}. \label{subfig:airfoils_6}]{\includegraphics[width=0.4\linewidth]{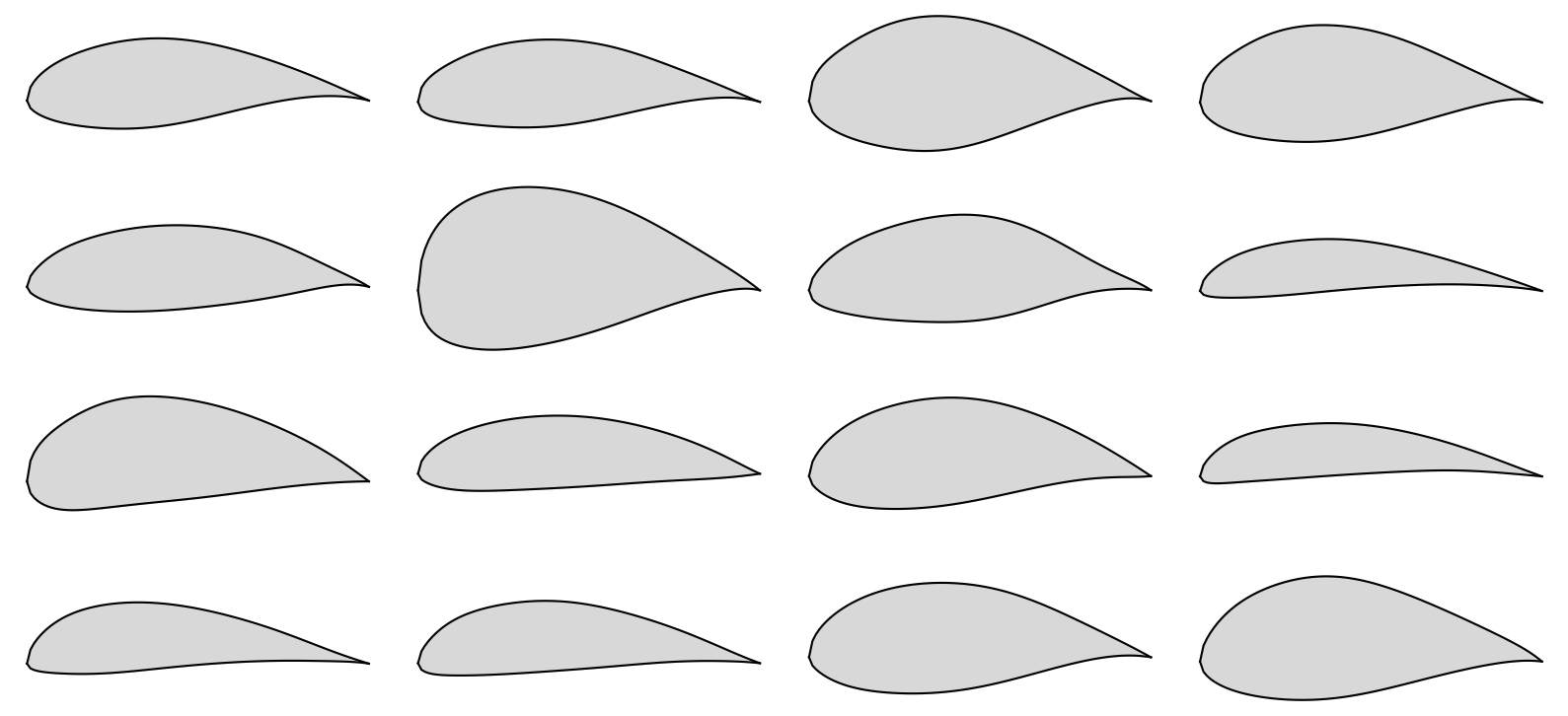}}
    }
    
    \caption{Random generated airfoils using energy-based approach with various settings and \textit{Dflow-SUR}}
    \label{fig:generatedAirfoil}
\end{figure}

We further visualize the generated airfoils in Figure~\ref{fig:generatedAirfoil} using both the energy-based approach with various $\lambda$ and $t_c$ settings and \textit{Dflow-SUR}. 

From the figure, we observe that using overly small or prematurely applied hyperparameters (e.g., $\lambda = 0.1$ or $t_c = 0.0$) does not degrade sample diversity but significantly increases the failure rate and fails to enforce physical constraints. Conversely, employing excessively large or belated hyperparameters (e.g., $\lambda = 100$ or $t_c = 0.8$) guarantees high-quality, constraint-satisfying outputs but sharply limits generative diversity. The samples generated by \textit{Dflow-SUR}, on the other hand, can strike the right balance between diversity, constraint satisfaction, and output quality.

In summary, the energy-based method’s sensitivity to guidance strength makes reliable generation challenging. In contrast, \textit{Dflow-SUR} requires only random initial guesses, and its decoupled execution naturally ensures both physical validity and high geometric quality of the generated samples.

\subsection{3D wing inverse design case}
\label{sect:wingCase}
In this section, we employ the generative inverse model to explore high aerodynamic performance wing candidates for the three-dimensional wing shape design. It is worth mentioning that our generative model is not used to solve a design optimization problem in this case; instead, it is used to generate desired design candidates under given constraints. 
Referring to the American Institute of Aeronautics and Astronautics (AIAA) Aerodynamic Design Optimization Discussion Group (ADODG)\footnote{AIAA ADODG webpage: \url{https://sites.google.com/view/mcgill-computational-aerogroup/adodg} (last accessed on 26 July 2025).} case 4.1 and previous benchmark investigations~\cite{lyu2015aerodynamic,li2021adjoint}, we set the multi-dimensional constraints to the generative model to achieve lift-to-drag ratio $L/D = 21.8$, which is the highest performance achieved using adjoint CFD solver (specifically by a design with $C_{L} = 0.5$ and $C_{D} = 0.0229$). 

\begin{figure}[ht!]
    \centering
    \subfloat[Wing sample generated using LHS]{%
        \includegraphics[width=0.5\textwidth]{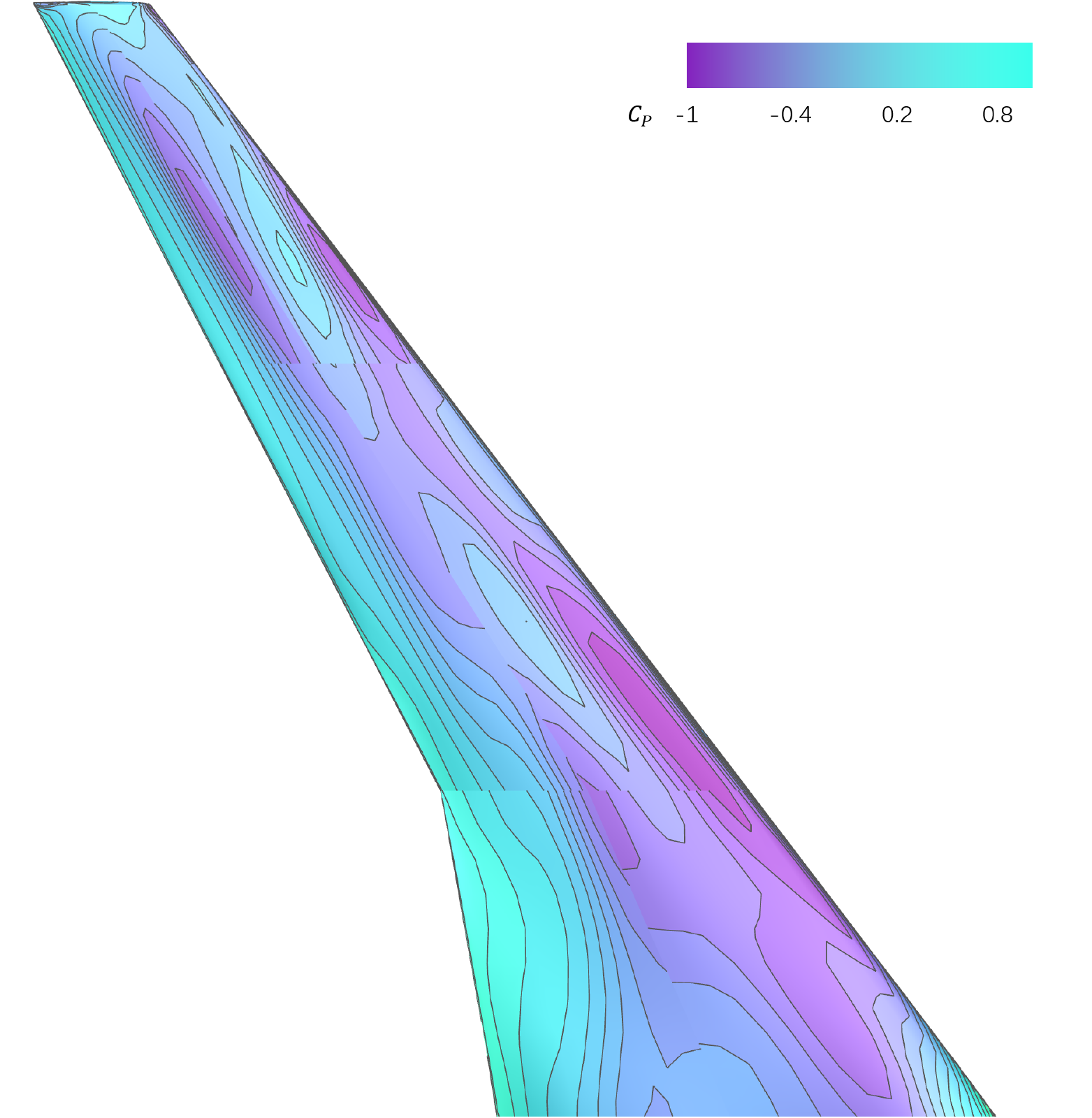}%
    }\hfill
    \subfloat[Wing sample generated using Dflow-SUR]{%
        \includegraphics[width=0.5\textwidth]{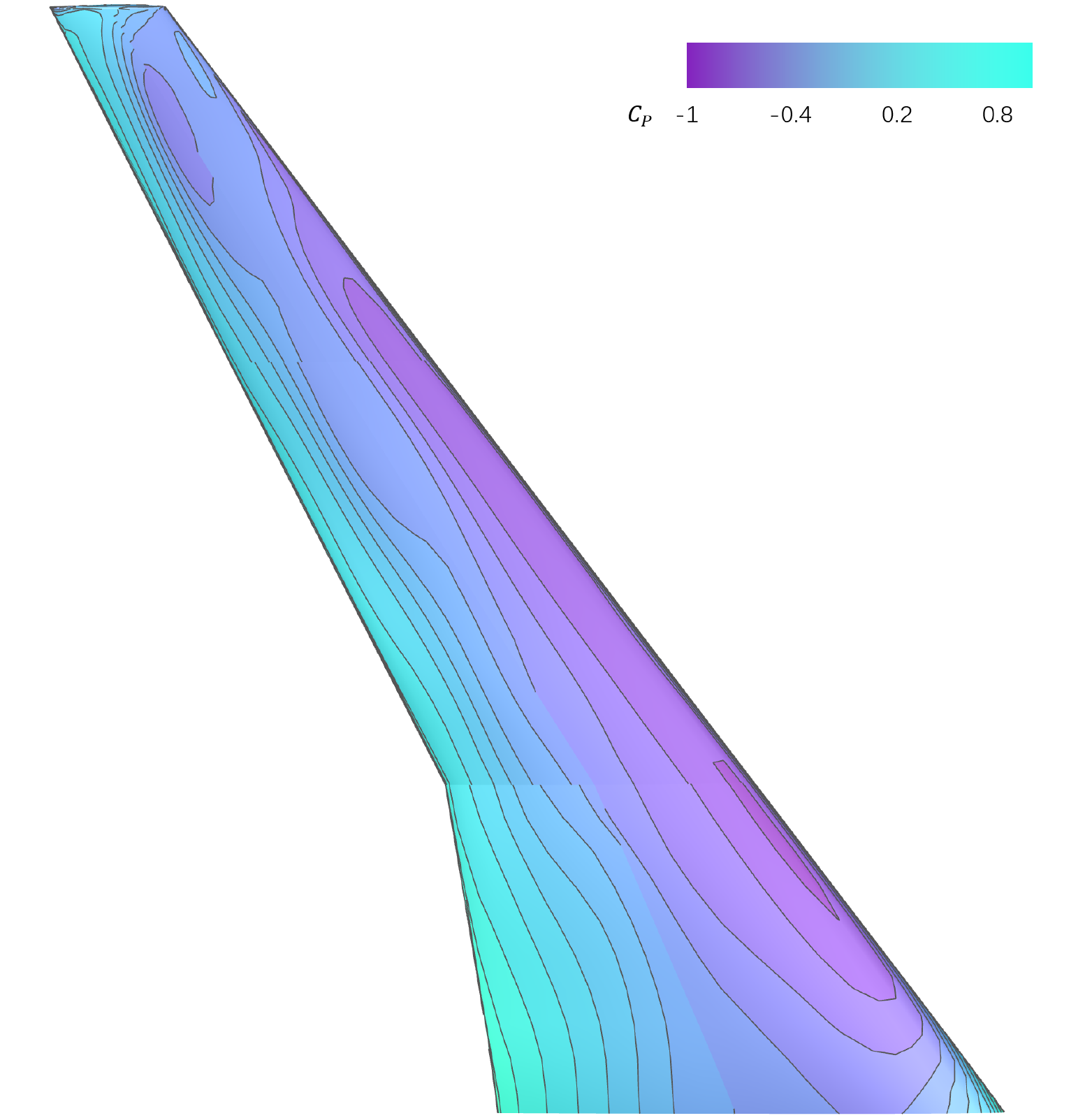}%
    }
    \caption{Wing sample $C_{P}$ distribution validated using ADflow.}
    \label{fig:wingAerodynamic}
\end{figure}

We further validate the wing samples using ADflow~\cite{Mader2020a,Kenway2019a}\footnote{ADflow repository: \url{https://github.com/mdolab/adflow.git} (last access on 26 July 2025)} CFD solver and visualize the pressure coefficient $C_{P}$ distributions of two samples as shown in Figure~\ref{fig:wingAerodynamic}. From the $C_P$ distributions, it is evident that, relative to conventional methods, \textit{Dflow-SUR} produces more aerodynamically-coherent shapes, yielding a more uniform $C_P$ variation from the leading edge to the trailing edge.

\begin{figure}[ht!]
    \centering
    \subfloat[Probability density plot]{%
        \includegraphics[width=0.8\textwidth]{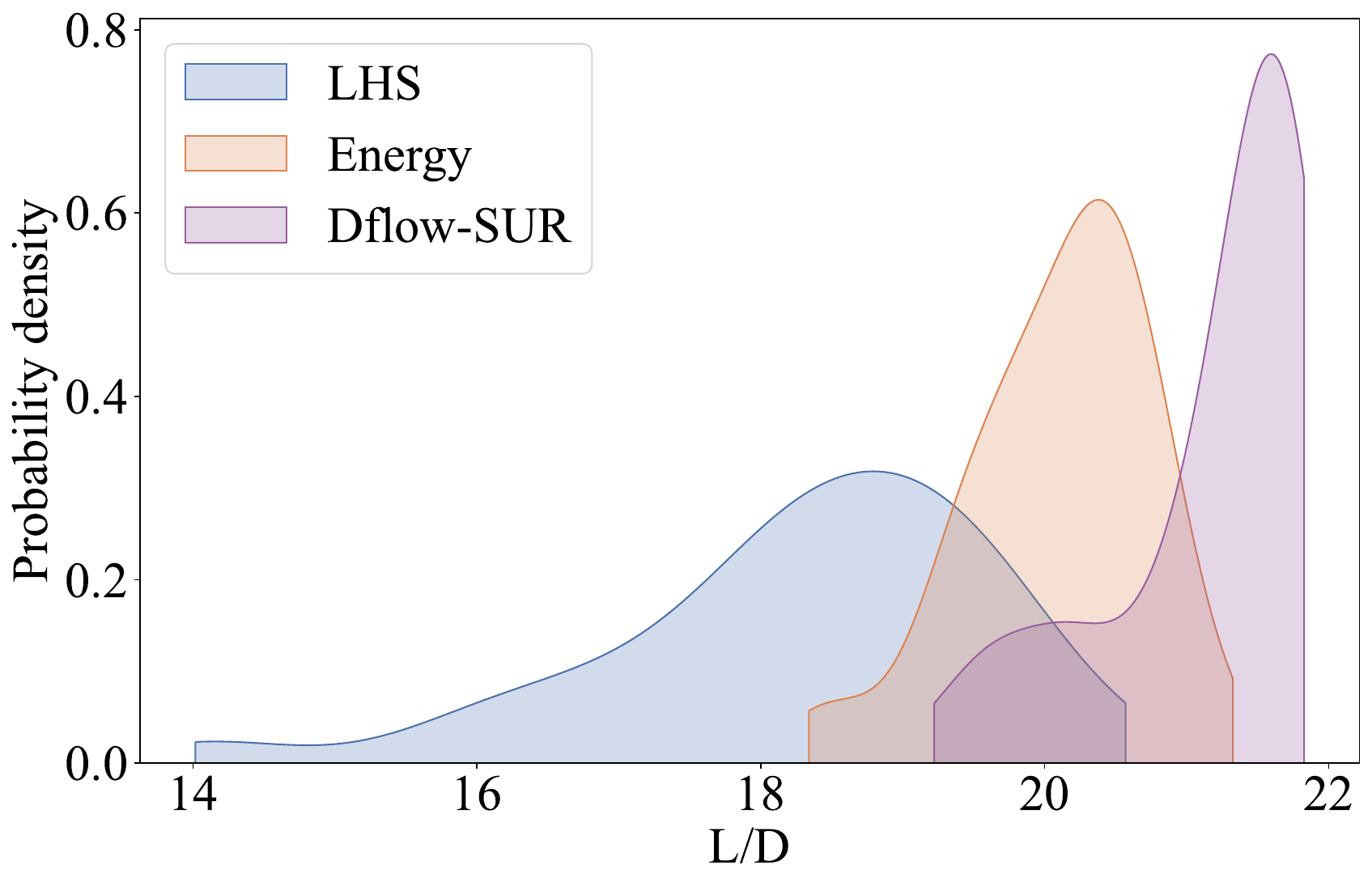}%
        \label{subfig:wing_PDF}
    }\\[1em]
    \subfloat[Violin plot]{%
        \includegraphics[width=0.8\textwidth]{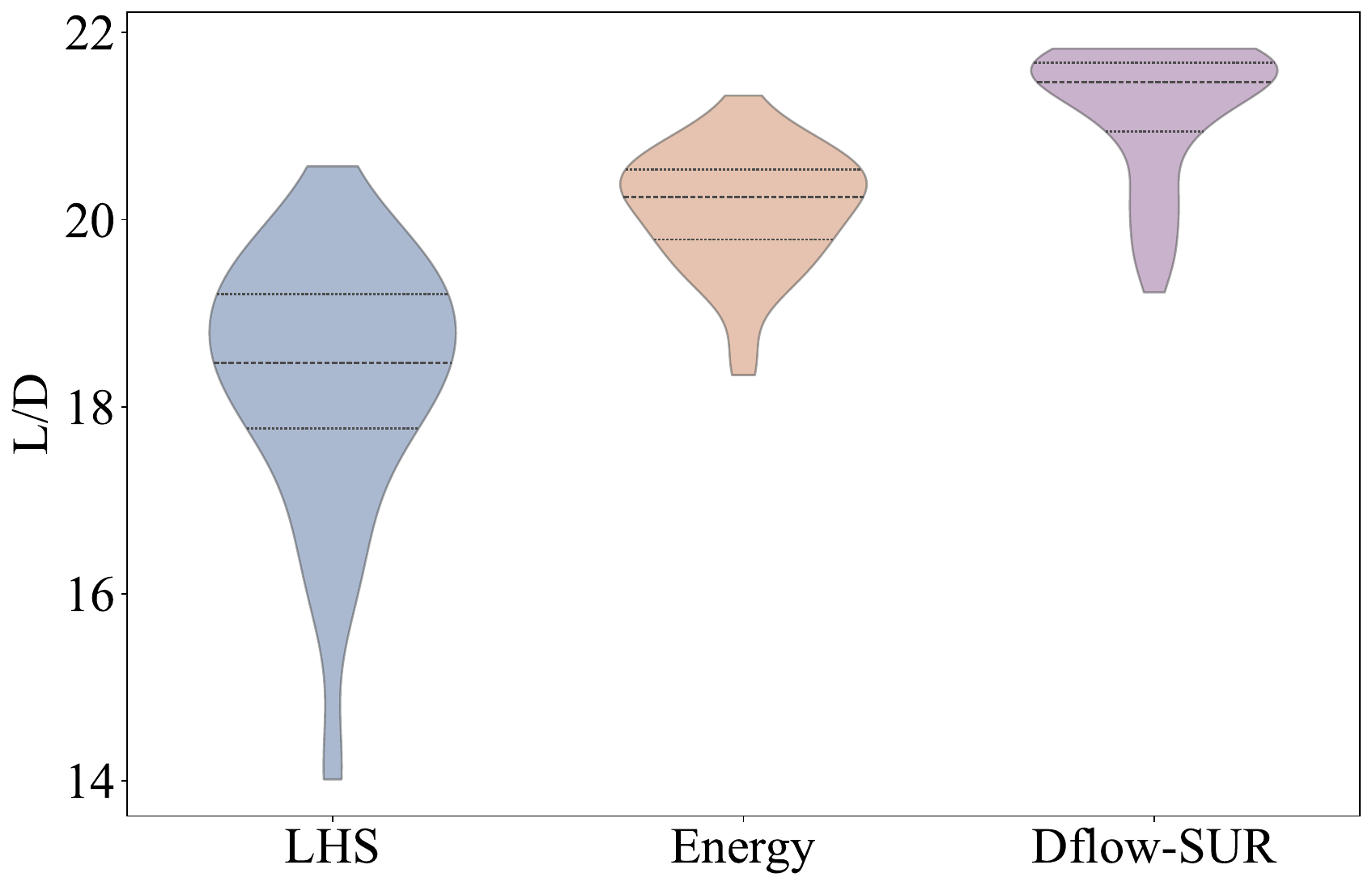}%
        \label{subfig:wing_violin}
    }\\[1em]

    \caption{Probability density and violin plot of wing samples $L/D$ distribution using three sampling approaches. (Three horizontal lines in each violin plot indicate the 25th percentile, median, and 75th percentile.)}
    \label{fig:wingPerformance}
\end{figure}

Figure~\ref{fig:wingPerformance} shows the probability density (Figure~\ref{subfig:wing_PDF}) and violin plot (Figure~\ref{subfig:wing_violin}) of wing samples' $L/D$ distributions using LHS, energy-based, and \textit{Dflow-SUR} approaches. The data shown herein are truncated at its observed minimum and maximum values. From the results, \textit{Dflow-SUR} outperforms both LHS and the energy-based approach as a sampling method: it achieves a higher mean $L/D$ ($21.1845$ as compared to $18.3998$ obtained using LHS and $19.8416$ from the energy-based approach) and a lower standard deviation ($0.7020$ as compared to $1.4641$ and $1.0201$ obtained using LHS and the energy-based approach, respectively). This indicates that \textit{Dflow-SUR} not only shifts the distribution toward higher aerodynamic performance but also concentrates design candidates more tightly around higher $L/D$ values, yielding a more efficient generation of high-performance geometries.

The wing samples and surrogate model are validated in a previous work by Li and Zhang~\cite{li2021data}, which were used to perform surrogate-based wing shape optimization of the Common Research Model (CRM) wing\footnote{NASA Common Research Model (CRM) webpage: \url{https://commonresearchmodel.larc.nasa.gov/} (last accessed on 26 July 2025).}. The wing samples are generated via Latin Hypercube Sampling (LHS) around the CRM wing baseline shape using a compact modal parameterization method, which is also based on another work by Li and Zhang~\cite{li2021adjoint}. The surrogate model on aerodynamic estimation is built with inputs of wing geometry $\mathbf{x}$, wing twists $\alpha_{twist}$, Mach number $M$, flight altitude $h$, angle of attack $\alpha$ and outputs $C_{L}$, $C_{D}$ and $C_{M}$. 

It is worth noting that this study focuses on investigating how to better integrate physical guidance into flow models. The performance of the resulting generative model depends on the accuracy of the surrogate model, which may lead to discrepancies with respect to CFD solutions. However, this issue pertains to the broader discussion between surrogate models and CFD, falling outside the scope of the present study.

\section{Conclusions}
\label{sect:conculsion}
In this paper, we presented \textit{Dflow‑SUR}, a data-driven approach suitable for physics-guided generative inverse design that decouples flow matching inference from physical loss optimization by differentiating throughout the entire generative trajectory. Unlike traditional conditional training and energy-based inference approaches, which suffer from gradient collisions and asynchronous dynamics, \textit{Dflow‑SUR} evaluates the physical loss only at the terminal sample and back-propagates its gradient to the initial noise input. 

We validate the effectiveness of the proposed approach on both 2D airfoil and 3D wing design tasks. In the airfoil case, \textit{Dflow‑SUR} improves generation accuracy by up to four orders of magnitude compared to the best-performing traditional method, while reducing generation time by $74.47\%$. As a high-performance aerodynamic sampler in the 3D wing case, \textit{Dflow‑SUR} generates a more concentrated and elevated distribution under prescribed aerodynamic constraints, achieving an $11.8\%$ increase in mean $L/D$ over Latin Hypercube Sampling and a $6.5\%$ improvement over the energy-based approach. As a result, the $C_P$ distributions of the generated samples indicate that \textit{Dflow‑SUR} produces more aerodynamically reasonable configurations.

In conclusion, \textit{Dflow‑SUR} framework delivers three principal benefits for physics‑informed generative inverse design, as demonstrated through a series of comparative experiments. First, it offers \textbf{superior guidance controllability}; by separating the flow‑matching inference from physical‑loss optimization, \textit{Dflow‑SUR} eliminates the gradient collisions inherent to tightly coupled methods, enabling more thorough generative and physical‑optimization processes. Second, it provides \textbf{uncertainty control in early denoising}. In this context, \textit{Dflow‑SUR} evaluates only the final design $\mathbf{x}_1$ and back‑propagates its physical gradient to the initial noise $\mathbf{x}_0$ via the chain rule, thereby avoiding the surrogate-model uncertainties that arise from out-of-distribution intermediates. Third, \textbf{hyperparameter robustness}; unlike energy‑based approaches that require meticulous tuning of coefficients and cutoff times, \textit{Dflow‑SUR} operates with essentially no additional hyperparameters, facilitating straightforward batch deployment.

To recap, the primary objective of \textit{Dflow‑SUR} is to enable the incorporation of physical information into the generative process with higher accuracy and efficiency, rather than to replace traditional high-fidelity CFD-based design optimization paradigms. The generative model is intended to explore physically reasonable conceptual designs, serving as a front-end tool for design space exploration. Given the substantial computational cost associated with physics-based evaluation in computational mechanics—unlike typical AI tasks—\textit{Dflow‑SUR}'s decoupled framework, which separates physical loss from flow matching and restricts physics evaluation to plausible designs, offers a practical and scalable solution.
\appendix
\section{Energy-based approach algorithm}
\label{sect:App_energyBasedAlg}

The pseudo code of the energy-based physics injection flow matching algorithm involving full and intermediate trajectory injection strategies is described in Algorithm~\ref{alg:energyAlg}.

\begin{algorithm}[H]
\caption{Energy-based Guided Flow Matching}
\begin{algorithmic}[1]
\Require Pretrained velocity field $u_t^\theta$, energy $\mathcal{E}(\cdot)$, coefficient $\lambda$, total steps $T$, cutoff time $t_c\in[0,1]$
\State Set $\Delta t = 1/T$
\State Initialize $\mathbf{x}_0 \sim \mathcal{N}(0,I)$
\For{$i = 0$ to $T-1$}
    \State $t_i = i\,\Delta t$
    \If{$t_i < t_c$}  
        \State $\dot{\mathbf{x}} \leftarrow u_{t_i}^\theta(\mathbf{x}_i, t_i)$
    \Else
        \State $\dot{\mathbf{x}} \leftarrow u_{t_i}^\theta(\mathbf{x}_i, t_i)\;-\;\lambda\,\nabla_{\mathbf{x}_i}\mathcal{E}(\mathbf{x}_i)$
    \EndIf
    \State $\mathbf{x}_{i+1} \leftarrow \mathbf{x}_i + \Delta t\,\dot{\mathbf{x}}$
\EndFor
\State \Return $\mathbf{x}_1$
\end{algorithmic}
\label{alg:energyAlg}
\end{algorithm}

\section{Airfoil case}
\subsection{Model performance}
\label{subsect:modelPerformance}
We present the model performance comparative study results in Table~\ref{tab:modelPerformanceResults}. We primarily compare the final reduced values of the physical loss $\mathcal{L}_{\mathrm{phys}}$, the achieved $C_{L}$ accuracy, and the computation time. To ensure a fair comparison involving the surrogate model, all experiments are conducted on CPUs. In summary, \textit{Dflow-SUR} demonstrates a four orders-of-magnitude improvement in physical loss and achieves an approximate $74.47\%$ reduction in runtime compared to the energy-based baseline (with $t_c = 0.6$ and $T = 1000$), highlighting its exceptional capability to learn and enforce physical constraints.

\begin{table}[!htb]
  \centering
  \renewcommand{\arraystretch}{1.2}
  \begin{tabular}{lcccc}
    \toprule
    \textbf{Method} & \textbf{T} & \textbf{$\mathcal{L}_{\mathrm{phys}}$} & \textbf{$C_L$} & \textbf{Time cost (s)} \\
    \midrule
    \multirow{1}{*}{Conditional training}
      & 2000  & $(5.35 \pm 0.86)\times10^{-3}$ & $0.627 \pm 0.028$ & 12779.44 \\
    \midrule
    \multirow{3}{*}{Energy-based ($t_c=0.0$)}
      & 200   & 0.1766         & 0.1957           & 6678.79 \\
      & 1000  & $(1.07 \pm 0.60)\times10^{-2}$ & $0.655 \pm 0.023$ & 3169.04 \\
      & 2000  & $(4.69 \pm 0.86)\times10^{-2}$ & $0.553 \pm 0.026$ & 6779.44 \\
    \midrule
    \multirow{3}{*}{Energy-based ($t_c=0.2$)}
      & 200   & 0.0927         & 0.1373           & 5395.10 \\
      & 1000  & $(1.51 \pm 0.79)\times10^{-3}$ & $0.683 \pm 0.008$ & 3192.07 \\
      & 2000  & $(1.22 \pm 0.10)\times10^{-2}$ & $0.596 \pm 0.005$ & 6002.10 \\
    \midrule
    \multirow{3}{*}{Energy-based ($t_c=0.6$)}
      & 200   & 0.0716         & 0.1407           & 3342.80 \\
      & 1000  & $(4.80 \pm 6.91)\times10^{-4}$ & $0.710 \pm 0.0045$ & 3136.75 \\
      & 2000  & $(4.24 \pm 0.81)\times10^{-3}$ & $0.639 \pm 0.006$ & 6485.92 \\
    \midrule
    \multirow{3}{*}{Energy-based ($t_c=0.8$)}
      & 200   & 0.0361         & 0.1748           & 1186.29 \\
      & 1000  & $(8.40 \pm 4.0)\times10^{-4}$  & $0.721 \pm 0.004$ & 3015.18 \\
      & 2000  & $(9.54 \pm 1.20)\times10^{-3}$ & $0.703 \pm 0.017$ & 6440.14 \\
    \midrule
    \textbf{Dflow-SUR} & \textbf{50} & \boldmath{$(4.80 \pm 6.91)\times10^{-8}$} & \boldmath{$0.699 \pm 6\times10^{-9}$} & \textbf{801} \\
    \bottomrule
  \end{tabular}
  \caption{Model performance comparison across methods and inference time steps ($T$).}
  \label{tab:modelPerformanceResults}
\end{table}

\subsection{Gradient alignment score}
\begin{figure}[htbp]
    \centering
    \subfloat[Physics injection when $t_c=0.2$ \label{subfig:uqEnergy_1}]{%
        \includegraphics[width=0.7\textwidth]{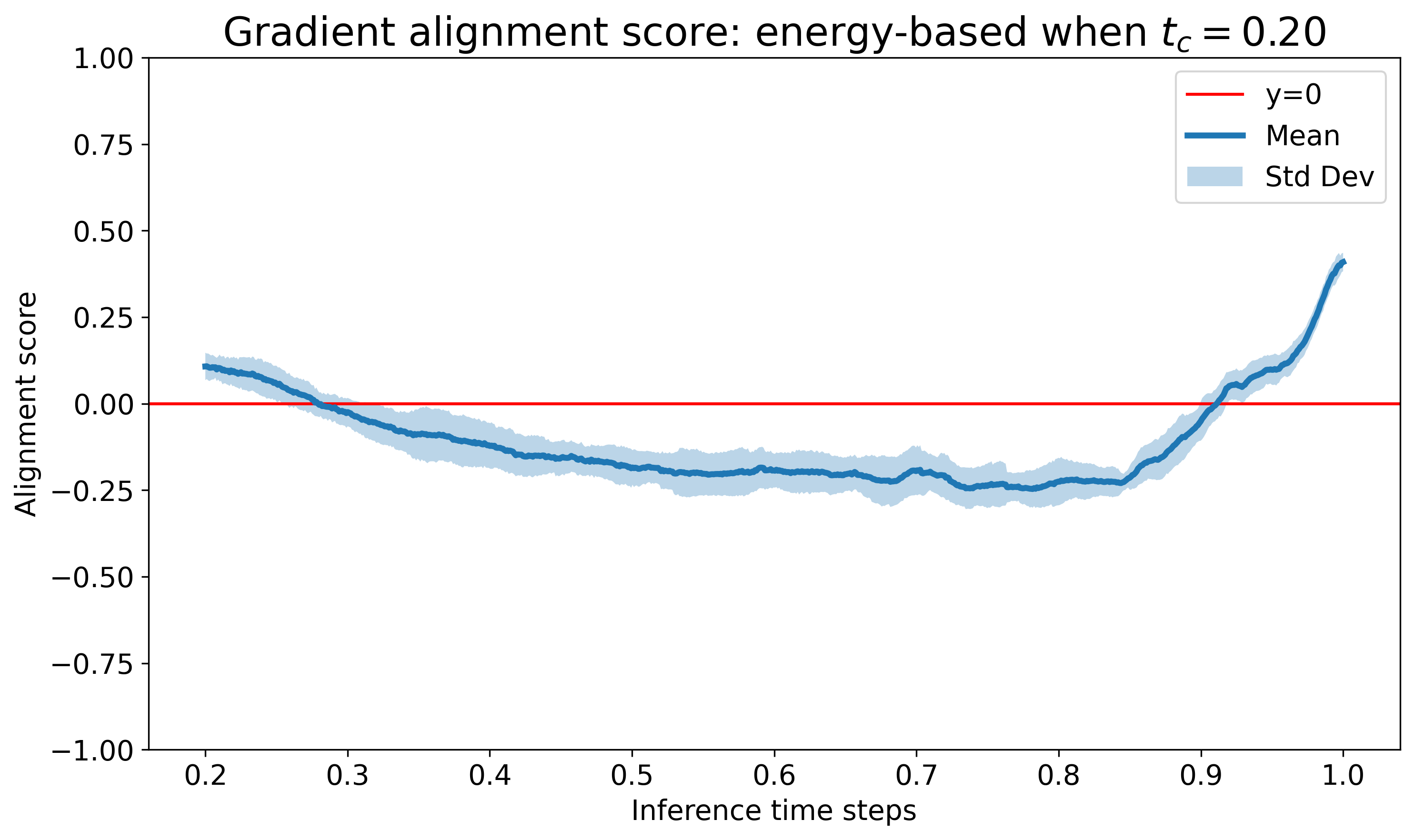}%
    }\quad
    \subfloat[Physics injection when $t_c=0.6$\label{subfig:uqEnergy_2}]{%
        \includegraphics[width=0.7\textwidth]{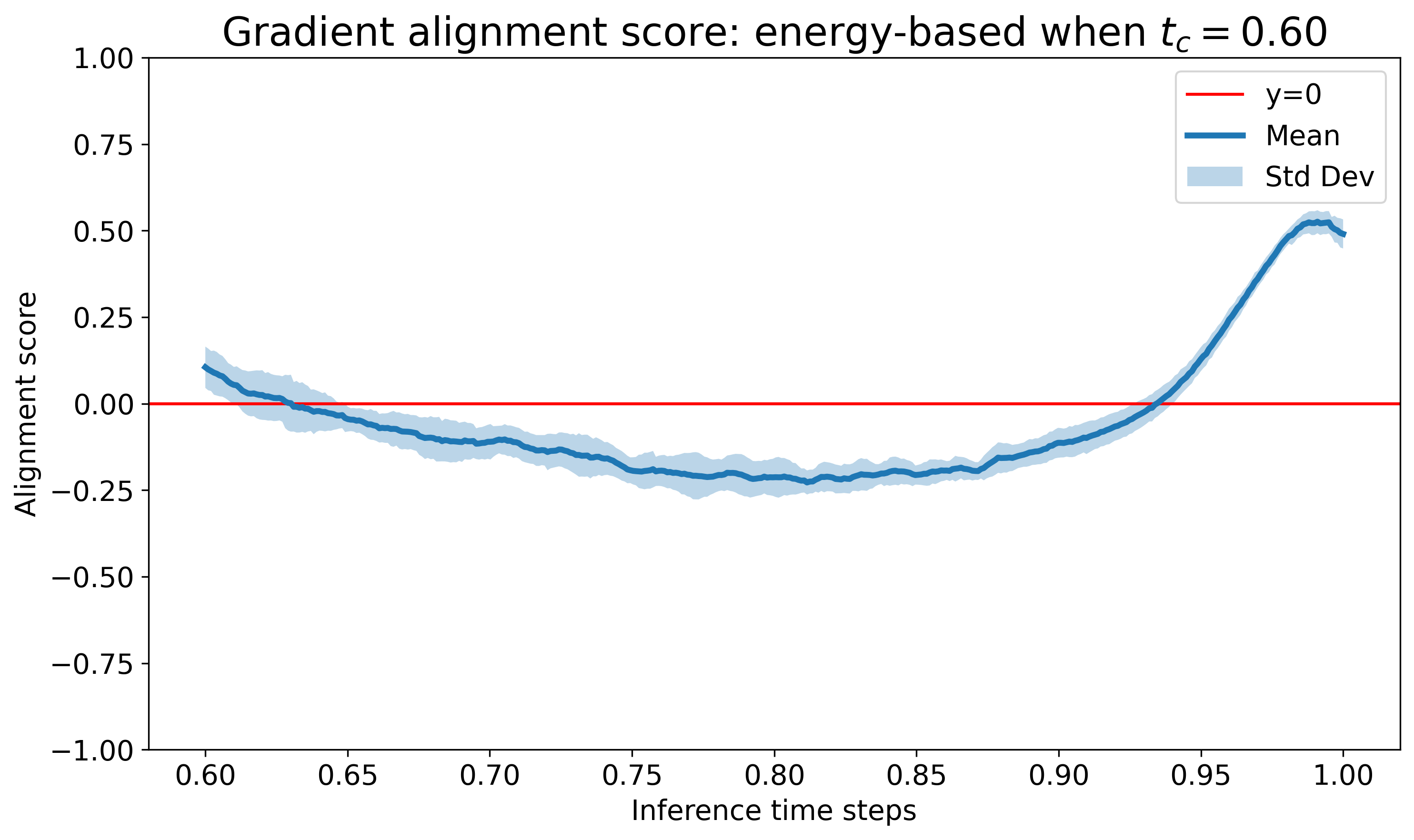}%
    }\\[1em]
    \subfloat[Physics injection when $t_c=0.8$\label{subfig:uqEnergy_3}]{%
        \includegraphics[width=0.7\textwidth]{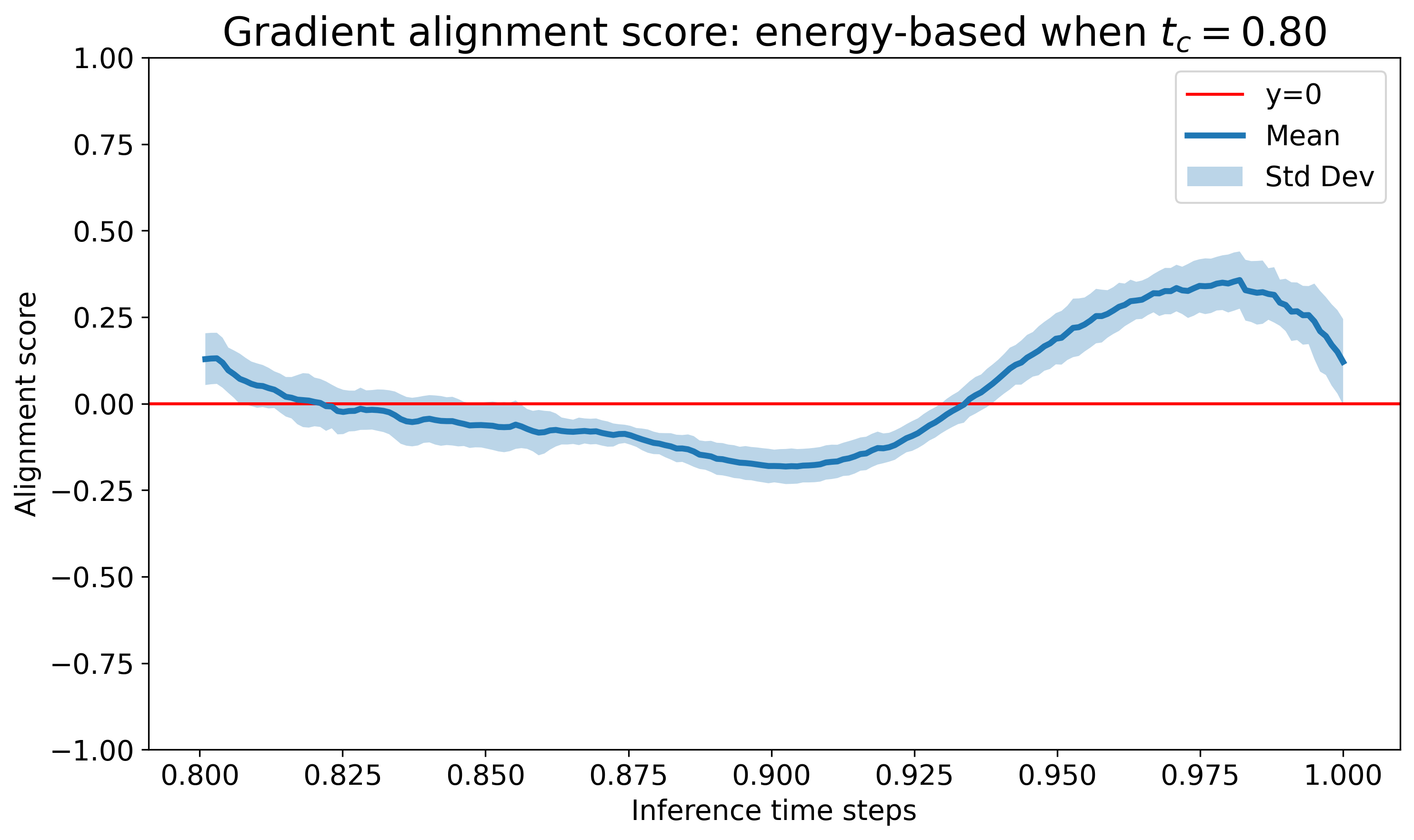}%
    }\quad
    \caption{Gradient alignment scores during the $\mathcal{L}_{\mathrm{phys}}$ injection phase of the energy-based inference ($T = 1000$, $\lambda = 10$) with varying $t_c$. The $x$-axis is aligned such that 0 indicates the onset of $\mathcal{L}_{\mathrm{phys}}$ injection.}
    \label{fig:alignmentScore_app}
\end{figure}

We further present the gradient alignment score of the energy-based approach when $t_c = 0.20, 0.60, 0.80$ (introduced in Section~\ref{subsect:physicalloss}) in Figure~\ref{fig:alignmentScore_app}. It can be observed that the gradient collision phenomenon persists during the primary inference phase. This indicates that the physical loss and flow matching loss consistently compete in their influence on design $\textbf{x}$, a behavior attributable to the issues arising from the injection of two guidance couplings.

\label{sect:gradientAlignmentScore}

\subsection{UQ of energy-based approach}
\label{sect:AppendixUncertainty}
\begin{figure}[htbp]
    \centering
    \subfloat[Physics injection when $t_c=0.0$ \label{subfig:uqEnergy_1}]{%
        \includegraphics[width=0.7\textwidth]{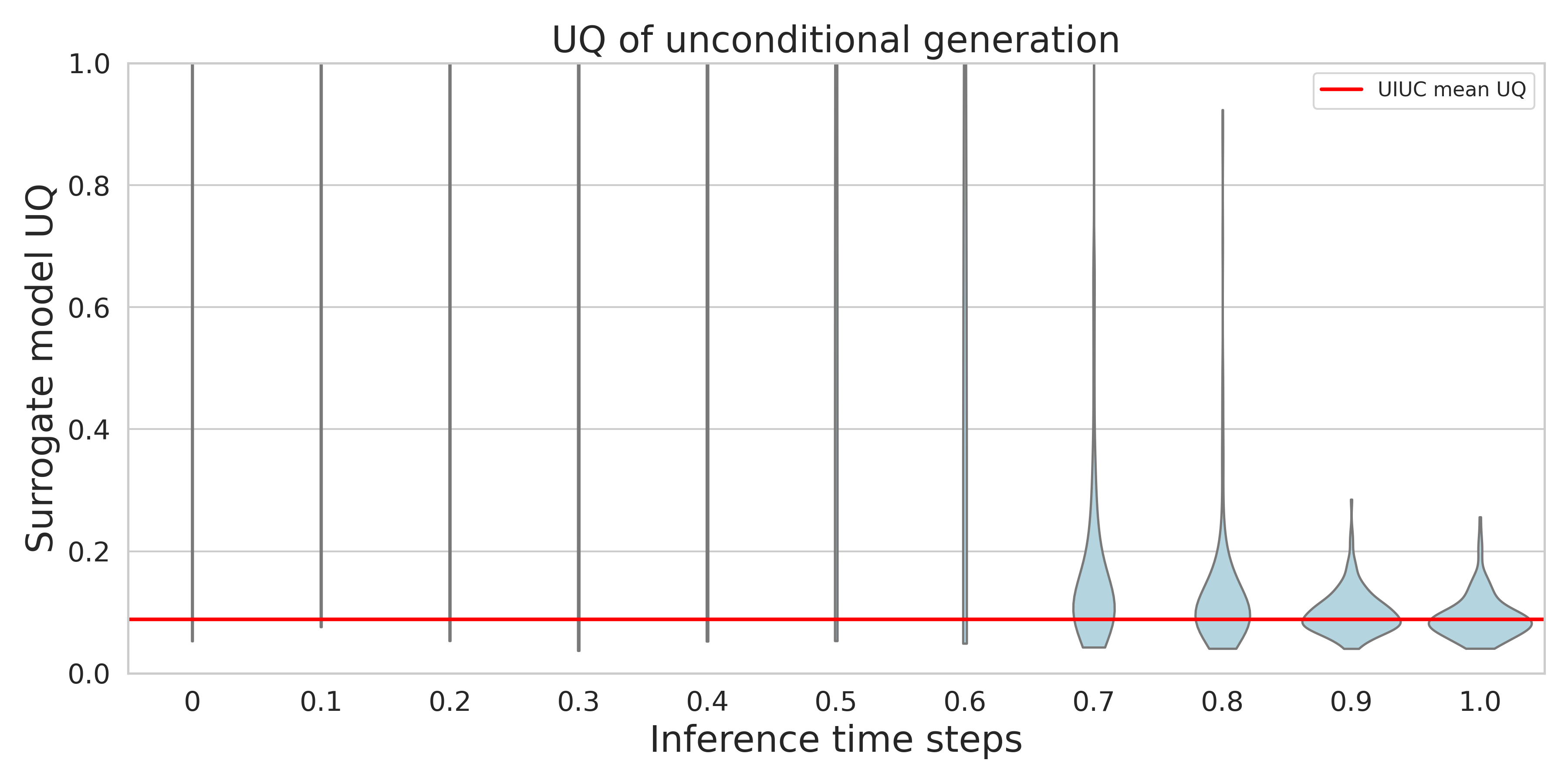}%
    }\quad
    \subfloat[Physics injection when $t_c=0.2$\label{subfig:uqEnergy_2}]{%
        \includegraphics[width=0.7\textwidth]{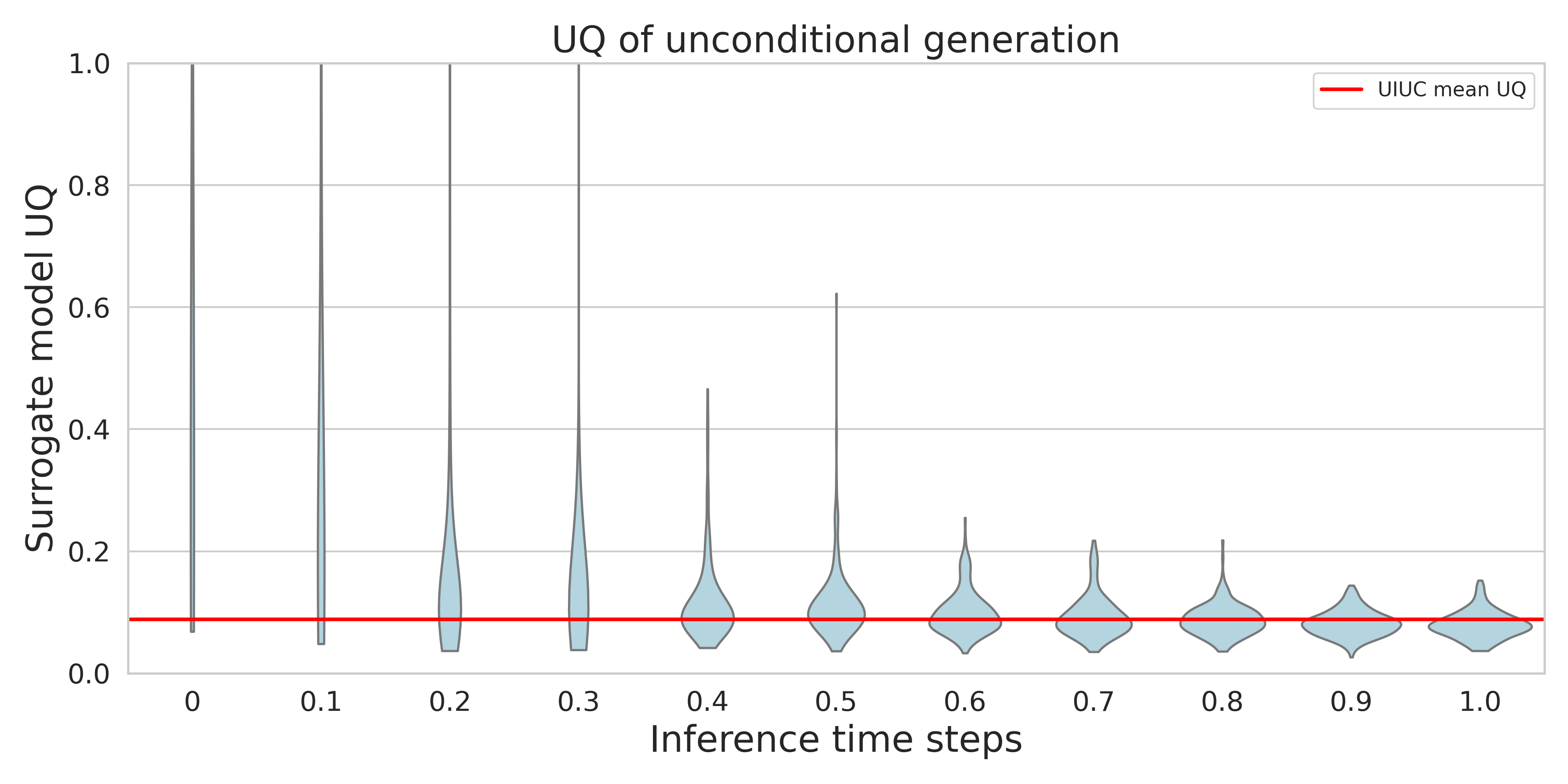}%
    }\\[1em]
    \subfloat[Physics injection when $t_c=0.6$\label{subfig:uqEnergy_3}]{%
        \includegraphics[width=0.7\textwidth]{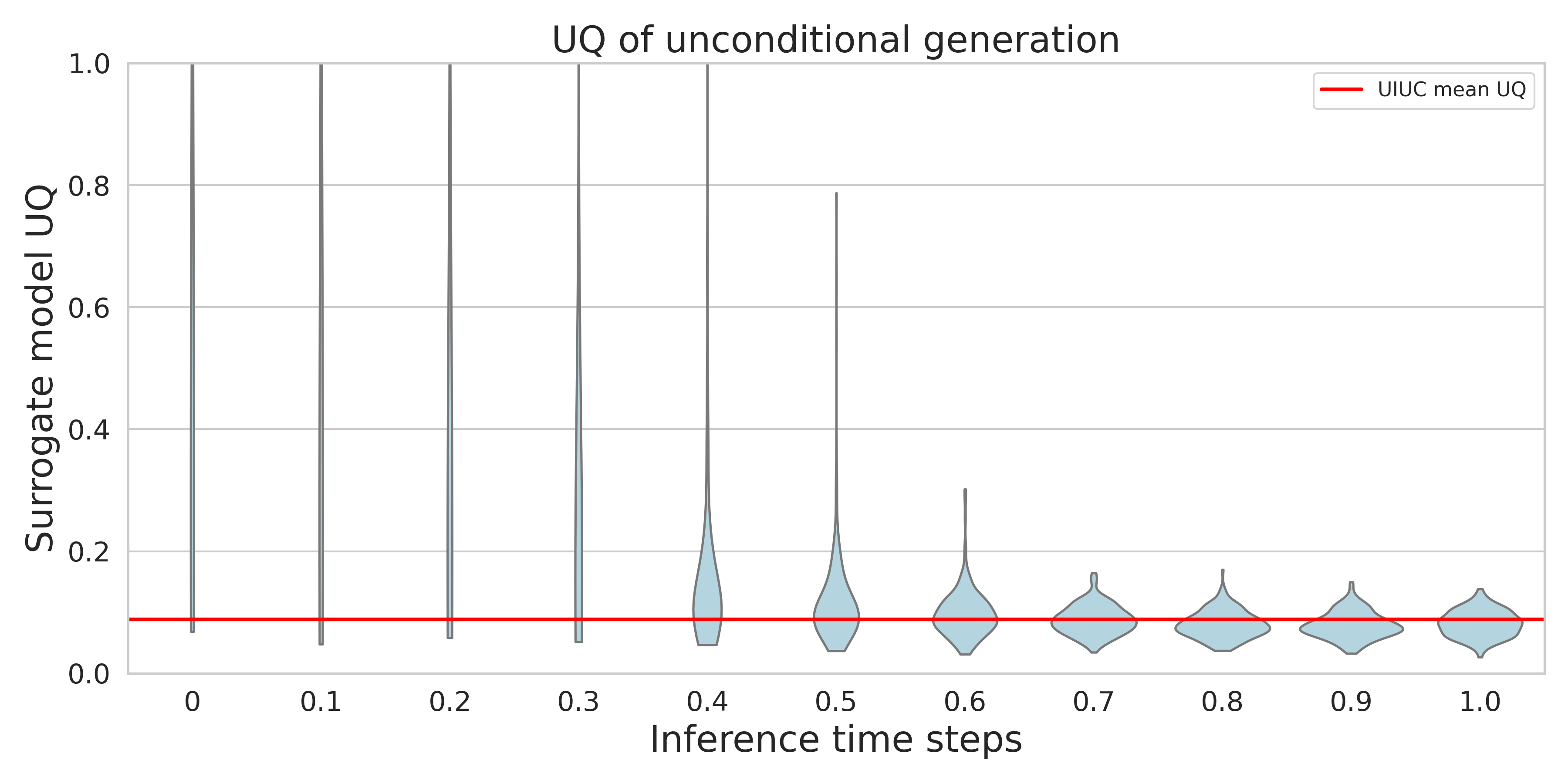}%
    }\quad
    \subfloat[Physics injection when $t_c=0.8$\label{subfig:uqEnergy_4}]{%
        \includegraphics[width=0.7\textwidth]{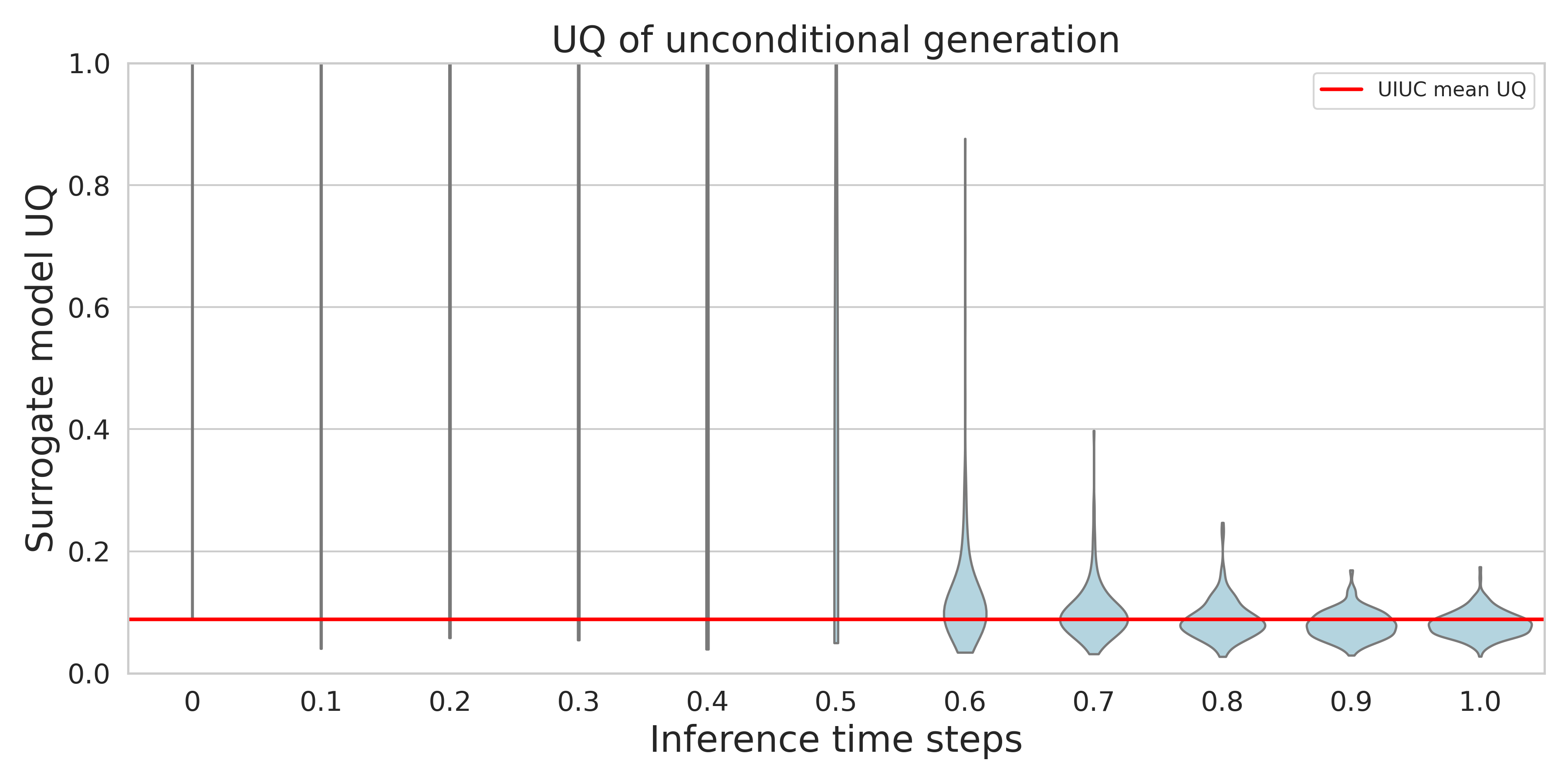}%
    }
    \caption{UQ of generative samples for energy-based approach (total inference time step = 1000) with various $T$ physics injection (the red line represents the UIUC data UQ mean).}
    \label{fig:uqEnergy}
\end{figure}

We further present the UQ of generated samples from the energy-based approach evaluated by the surrogate model at each inference time step in Figure~\ref{fig:uqEnergy}. From analyzing the figures, we can observe that samples generated by the energy-based method during its generation process still exhibit high surrogate-evaluated UQ in the inference phase, thereby misleading the loss guidance.

\bibliographystyle{unsrt}  

\clearpage
\bibliography{main_ref}

\end{document}